\documentclass[12pt]{article}

\usepackage[dvipsname]{xcolor}

\usepackage[dvipdfmx]{graphicx}
\usepackage{graphicx}
\usepackage{amsmath}        
\usepackage{amssymb}        
\usepackage{slashed}
\usepackage{bm}

\def\mydate{June 17, 2020 \hfill OU-HET 1043}
\def\ignore#1{{}}

\def\go{\rightarrow}
\def\dd{\partial}
\def\tr{{\rm tr}\,}

\def\eff{{\rm eff}}
\def\SM{{\rm SM}}
\def\KK{{\rm KK}}
\def\EM{{\rm EM}}
\def\onehalf{\hbox{$\frac{1}{2}$}}
\def\la{\langle}
\def\ra{\rangle}

\def\app{{\rm app}}

\def\mybig{\displaystyle \strut }

\def\myfrac#1#2{\frac{\mybig #1}{\mybig #2}}

\def\mymat#1#2{\begin{matrix}#1 \cr \noalign{\kern -2pt} #2\end{matrix}}

\def\mynoalign{\noalign{\kern 4pt}}
\def\mysnoalign{\noalign{\kern 3pt}}

\makeatletter
\@addtoreset{equation}{section}
\makeatother

\begin{document}

\thispagestyle{empty}


\leftline{\mydate}
\rightline{KYUSHU-HET-209}

\vskip 4.0cm

\baselineskip=30pt plus 1pt minus 1pt

\begin{center}

{\LARGE \bf The effective potential  and universality}

{\LARGE \bf in GUT inspired gauge-Higgs unification}

\end{center}


\baselineskip=22pt plus 1pt minus 1pt

\vskip 2.0cm

\begin{center}
{\bf Shuichiro Funatsu$^1$, Hisaki Hatanaka$^2$, Yutaka Hosotani$^3$,}

{\bf Yuta Orikasa$^4$ and Naoki Yamatsu$^5$}

\baselineskip=17pt plus 1pt minus 1pt

\vskip 10pt
{\small \it $^1$Institute of Particle Physics and Key Laboratory of Quark and Lepton 
Physics (MOE), Central China Normal University, Wuhan, Hubei 430079, China} \\
{\small \it $^2$Osaka, Osaka 536-0014, Japan} \\
{\small \it $^3$Department of Physics, Osaka University, 
Toyonaka, Osaka 560-0043, Japan} \\
{\small \it $^4$Institute of Experimental and Applied Physics, Czech Technical University in Prague,} \\
{\small \it Husova 240/5, 110 00 Prague 1, Czech Republic} \\
{\small \it $^5$Department of Physics, Kyushu University, Fukuoka 819-0395, Japan} \\

\end{center}

\vskip 2.0
cm
\baselineskip=18pt plus 1pt minus 1pt

\begin{abstract}
The effective potential  for the Aharonov-Bohm phase $\theta_H$  in the fifth dimension
in GUT inspired $SO(5)\times U(1) \times SU(3)$ gauge-Higgs unification is evaluated
to show that dynamical electroweak symmetry breaking takes place with $\theta_H \not= 0$, 
the 4D Higgs boson mass 125$\,$GeV  being generated at the quantum level. 
The cubic and quartic self-couplings $(\lambda_3, \lambda_4)$ of the Higgs boson are 
found to satisfy universal  relations, i.e.\ they are determined, to high accuracy,  solely by $\theta_H$,
irrespective of values of other parameters in the model.  For $\theta_H=0.1$ ($0.15$), 
$\lambda_3$ and $\lambda_4$ are smaller than those in the standard model by 7.7\% (8.1\%) 
and 30\% (32\%),  respectively.
\end{abstract}

\newpage

\baselineskip=20pt plus 1pt minus 1pt
\parskip=0pt

\section{Introduction}

The Higgs boson is responsible for the electroweak symmetry breaking in the
standard model (SM).
The Higgs potential is arranged such that the Higgs field spontaneously develops a nonvanishing
vacuum expectation value.  Its couplings to quarks and leptons (Yukawa couplings)
are determined such that the observed quark-lepton mass spectrum is reproduced.
Although the SM seems consistent with almost all experimental data so far obtained, 
it is yet to be seen whether or not  the Higgs boson is exactly what is postulated in 
the SM.  Unlike the gauge sector in the SM, the Higgs sector lacks a principle, which
leaves arbitrariness in the theory.  
The Higgs boson  mass acquires large quantum corrections which must be cancelled
by fine-tuning of  parameters in the model.

One approach to overcome these difficulties is  gauge-Higgs unification in which
the 4D Higgs boson is identified with the 4D fluctuation mode of an Aharonov-Bohm (AB) phase
in the fifth dimension.  The 4D Higgs field is contained in the extra-dimensional
component of gauge potentials.
As an AB phase the Higgs boson is massless at the tree level, but acquires 
a finite mass at the quantum level, independent of a cutoff scale and regularization method.
The gauge hierarchy problem is naturally solved.\cite{Hosotani1983}-\cite{Kubo2002}

Recently substantial advances have been made in gauge-Higgs unification.
Realistic models have been constructed which yield nearly the same phenomenology
as the SM at low energies and give many predictions to be explored at LHC and ILC.
Most of gauge-Higgs unification models are constructed on orbifolds such as 
$M^4 \times (S^1/Z_2)$ and the Randall-Sundrum (RS) warped space.
Chiral fermions naturally emerge on orbifolds.\cite{Pomarol1998}
The $SU(2)_L$ doublet Higgs field must appear as a zero mode of the fifth-dimensional 
component of gauge fields.  This condition leads to gauge groups such as
$SU(3) \times U(1)_X \times SU(3)_C$ or $SO(5) \times U(1)_X \times SU(3)_C$, 
among which the latter accommodates the custodial symmetry in the Higgs sector.
Quark-lepton multiplets are introduced such that with orbifold conditions specified
zero modes appear precisely for quarks and leptons, but not for exotic light fermions.
They must have observed couplings to $W$ and $Z$ bosons, and their masses
must be reproduced.  
Further the effective potential for the AB phase $\theta_H$ must have a global minimum
at $\theta_H \not= 0$ so that  the electroweak gauge symmetry is dynamically broken to 
$U(1)_\EM$.    As a model satisfying  these conditions  $SO(5) \times U(1)_X \times SU(3)_C$
gauge-Higgs unification is formulated in the RS space.\cite{Scrucca2003}-\cite{FHHOY2019b}

In the RS space, which is an AdS spacetime sandwiched by UV and IR branes, wave
functions of dominant components of $W$ and $Z$ bosons are almost constant in the bulk region
so that gauge couplings of quarks and leptons turn out nearly the same as those in the SM.
The hierarchy between the Kaluza-Klein (KK) mass scale ($\sim 10\,$TeV) and 
the weak scale ($\sim 100\,$GeV) naturally emerges.   
Two typical ways of  introducing fermions have been investigated.
In one type of the models (the A model) quarks and leptons are introduced in the vector 
representation of $SO(5)$.  The model predicts large parity violation in the $Z'$ couplings 
of quarks and leptons, which can be checked in the early stage of the ILC experiments
with  polarized electron and positron beams.\cite{FHHO2017ILC, Yoon2018b, Bilokin2017, Fujii2017}

It has been noticed, however, that there arises a difficulty in promoting the A model 
to grand unification.\cite{Burdman2003}-\cite{Englert2020}
The natural extension of the $SO(5) \times U(1)_X \times SU(3)_C$ model is
$SO(11)$ gauge-Higgs grand unification.\cite{HosotaniYamatsu2015}  
Up-type quarks are contained in the spinor
representation of $SO(11)$, but not in the vector representation so that 
up-type quarks in the A model do not appear from the $SO(11)$ gauge-Higgs unification.
A new way of introducing fermion multiplets has been found which can be embedded into
the $SO(11)$ gauge-Higgs grand unification.\cite{FHHOY2019a}  
In this GUT inspired model, or the B model, 
quarks and leptons are introduced in the spinor and singlet representations of $SO(5)$.
It has been shown that quarks and leptons have correct gauge couplings.  
Furthermore the flavor mixing is nicely incorporated with gauge-invariant brane
interactions in the B model.  The CKM matrix is obtained,  and remarkably 
flavor changing neutral current (FCNC) interactions are naturally suppressed.\cite{FHHOY2019b}

In this paper we evaluate the effective potential $V_\eff (\theta_H)$ in  
GUT inspired $SO(5) \times U(1)_X \times SU(3)_C$ gauge-Higgs unification.
It will be shown that with appropriate choice of parameters $V_\eff (\theta_H)$ 
has global minimum at $\theta_H \not= 0$ and the Higgs boson mass $m_H = 125\,$GeV
is obtained.
The cubic and quartic self-couplings of the Higgs boson are determined from 
$V_\eff (\theta_H)$.  We shall show that those cubic and quartic self-couplings are,
to high accuracy, determined as functions of $\theta_H$ only.  
They do not depend on other parameters of the theory.  
It will be explained how this universality results in the model.

The effective potential $V_\eff (\theta_H)$ is important in discussing phase transitions
at finite temperature as well.  Recently a possibility of having first-order phase transitions in 
gauge-Higgs unification has been argued.\cite{Adachi2019}  At the moment the nature of phase 
transitions at finite temperature in $SO(5) \times U(1) \times SU(3)$ gauge-Higgs unification
remains unclear.

The Higgs boson as an AB phase in gauge-Higgs unification has similarity to that in 
composite Higgs models in which the Higgs boson appears 
as a pseudo-Nambu-Goldstone boson.\cite{ACP2005, Giudice2007, compositeHiggsRev}
In both scenarios the Higgs boson field has a character of a phase, but has
a quite different mechanism for acquiring its mass.  In gauge-Higgs unification the Higgs 
boson mass is generated by gauge-invariant dynamics of the AB phase, whereas
it results from ungauged part of global symmetry in composite Higgs models.
Further in gauge-Higgs unification left-handed and right-handed components of quarks and leptons
are normally localized in opposite branes; if  left-handed components are localized near UV (IR)
brane, then right-handed components are localized near IR (UV) brane.
In typical composite Higgs models all light quarks and leptons are assumed to be localized near UV brane.
This leads to big difference in phenomenology associated with $Z'$  or techni-rho bosons.
In gauge-Higgs unification in RS space there appears large parity violation in $Z'$ couplings
of quarks and leptons,\cite{FHHO2017ILC, Funatsu:2014fda}  
whereas such asymmetry is absent in composite Higgs models.
Gauge-Higgs unification is strictly regulated by gauge principle.

The paper is organized as follows.  
In Section 2 the model is introduced.
In Section 3 the effective potential $V_\eff (\theta_H)$ is evaluated. 
We show that dynamical EW symmetry breaking takes place.
The cubic and quartic self-couplings, $\lambda_3$ and $\lambda_4$,  of the Higgs boson 
are evaluated  from $V_\eff (\theta_H)$.  It is observed there that $\lambda_3$ and $\lambda_4$
are determined to high accuracy  as functions of $\theta_H$, irrespective of other parameters 
in the model.  
The origin of the  $\theta_H$ universality in the RS space is clarified in Section 4.
In Section 5 the spectrum of dark fermions is evaluated. 
Section 6 is devoted to summary.
Mass spectra of all fields in the model are summarized in Appendix A.
Functions used for the evaluation of $V_\eff (\theta_H)$ are summarized in Appendix B.

\section{Model}

The GUT inspired $SO(5)\times U(1)_X \times SU(3)_C$ gauge-Higgs unification 
has been introduced in refs.\ \cite{FHHOY2019a, FHHOY2019b}. 
It is defined in the RS warped space with metric given by\cite{RS1999}
\begin{align}
ds^2= g_{MN} dx^M dx^N =
e^{-2\sigma(y)} \eta_{\mu\nu}dx^\mu dx^\nu+dy^2,
\label{Eq:5D-metric}
\end{align}
where $M,N=0,1,2,3,5$, $\mu,\nu=0,1,2,3$, $y=x^5$,
$\eta_{\mu\nu}=\mbox{diag}(-1,+1,+1,+1)$,
$\sigma(y)=\sigma(y+ 2L)=\sigma(-y)$,
and $\sigma(y)=ky$ for $0 \le y \le L$.
In terms of the conformal coordinate $z=e^{ky}$
($1\leq z\leq z_L=e^{kL}$) in the region $0 \leq y \leq L$ 
\begin{align}
ds^2=  \frac{1}{z^2} \bigg(\eta_{\mu\nu}dx^{\mu} dx^{\nu} + \frac{dz^2}{k^2}\bigg) .
\label{Eq:5D-metric-2}
\end{align}
The bulk region $0<y<L$ ($1<z<z_L$) is anti-de Sitter (AdS) spacetime 
with a cosmological constant $\Lambda=-6k^2$, which is sandwiched by the
UV brane at $y=0$ ($z=1$) and the IR brane at $y=L$ ($z=z_L$).  
The KK mass scale is $m_{\rm KK}=\pi k/(z_L-1) \simeq \pi kz_L^{-1}$
for $z_L\gg 1$.

In addition to gauge fields $A_M^{SU(3)_C}$, $A_M^{SO(5)}$ and $A_M^{U(1)_X}$
of $SU(3)_C$,  $SO(5)$, and $U(1)_X$, we introduce matter fields listed in Table \ref{Tab:matter}.
Fields defined in the bulk satisfy orbifold boundary conditions.
Each gauge field satisfies
\begin{align}
&\begin{pmatrix} A_\mu \cr  A_{y} \end{pmatrix} (x,y_j-y) =
P_{j} \begin{pmatrix} A_\mu \cr  - A_{y} \end{pmatrix} (x,y_j+y)P_{j}^{-1}
\label{BC-gauge}
\end{align}
where $(y_0, y_1) = (0, L)$. 
$P_0=P_1= I_3$ for  $A_M^{SU(3)_C}$  and  $P_0=P_1= 1$ for $A_M^{U(1)_X}$.  
$P_0=P_1 = P_{\bf 5}^{SO(5)} = {\rm diag}\,  (I_4, - 1)$ for $A_M^{SO(5)}$ in the vector
representation and 
$P_0=P_1 =P_{\bf 4}^{SO(5)} = {\rm diag}\,  (I_2, -I_2)$ in the spinor representation, 
respectively.
Quark and lepton multiplets satisfiy
\begin{align}
&\Psi_{({\bf 3,4})}^{\alpha} (x, y_j - y) = 
- P_{\bf 4}^{SO(5)} \gamma^5 \Psi_{({\bf 3,4})}^{\alpha} (x, y_j + y) ~, \cr
&\Psi_{({\bf 3,1})}^{\pm \alpha}  (x, y_j - y) =
\mp \gamma^5 \Psi_{({\bf 3,1})}^{\pm \alpha}  (x, y_j + y) ~, \cr
&\Psi_{({\bf 1,4})}^{\alpha} (x, y_j - y) = 
- P_{\bf 4}^{SO(5)} \gamma^5 \Psi_{({\bf 1,4})}^{\alpha} (x, y_j + y) ~,
\label{quarkleptonBC1}
\end{align}
where $\alpha = 1\sim 3$.
Dark fermion multiplets satisfiy
\begin{align}
&\Psi_F^\beta (x, y_j - y) = 
(-1)^j  P_{\bf 4}^{SO(5)} \gamma^5 \Psi_F^\beta (x, y_j + y) ~, \cr
&\Psi_{({\bf 1,5})}^{\pm \gamma} (x, y_j - y) = 
\pm P_{\bf 5}^{SO(5)} \gamma^5 \Psi_{({\bf 1,5})}^{\pm \gamma} (x, y_j + y) ~,
\label{darkFBC1}
\end{align}
where $\beta = 1 \sim N_F$ and $\gamma = 1 \sim N_V$.

\begin{table}[bht]
{
\renewcommand{\arraystretch}{1.2}
\begin{center}
\caption{$SU(3)_C\times SO(5) \times U(1)_X$ content of matter fields is shown
in the GUT inspired B model and previous A model.
In the A model only $SU(3)_C\times SO(4) \times U(1)_X$
symmetry is preserved on the UV brane so that the $SU(2)_L \times SU(2)_R$ content
is shown for brane fields. The B model is analyzed in the present paper. 
}
\vskip 10pt
\begin{tabular}{|c||c|c|}
\hline
 & {B model} & {A model}
 \\ \hline \hline
quark 
 & $\Psi_{({\bf 3}, {\bf 4})}^\alpha : ({\bf 3}, {\bf 4})_{\frac{1}{6}}$,  
    $\Psi_{({\bf 3}, {\bf 1})}^{\pm \alpha} : ({\bf 3}, {\bf 1})_{-\frac{1}{3}}^\pm$
 & $\Psi_1^\alpha : ({\bf 3}, {\bf 5})_{\frac{2}{3}}$, 
    $\Psi_2^\alpha : ({\bf 3}, {\bf 5})_{-\frac{1}{3}}$ 
\\ \hline
lepton
 & $\Psi_{({\bf 1}, {\bf 4})}^\alpha : \strut ({\bf 1}, {\bf 4})_{-\frac{1}{2}}$ 
 & $\Psi_3^\alpha : ({\bf 1}, {\bf 5})_{-1}$
    $\Psi_4^\alpha : ({\bf 1}, {\bf 5})_{0}$  
\\ \hline
dark fermion 
 & $\Psi_F^\beta : ({\bf 3}, {\bf 4})_{\frac{1}{6}}$, 
    $\Psi_{({\bf 1}, {\bf 5})}^{\pm \gamma} : ({\bf 1}, {\bf 5})_{0}^\pm$ 
 & $\Psi_F^\delta : ({\bf 1}, {\bf 4})_{\frac{1}{2}}$ 
\\ \hline \hline
brane fermion
 & $\chi^\alpha : ({\bf 1}, {\bf 1})_{0} $ 
 & $\begin{matrix} 
      \hat{\chi}^q_{1, 2, 3 R} : ({\bf 3}, [{\bf 2,1}])_{\frac{7}{6}, \frac{1}{6}, -\frac{5}{6}} 
 \cr
      \hat{\chi}^l_{1, 2, 3 R} : ({\bf 1}, [{\bf 2,1}])_{ -\frac{3}{2}, \frac{1}{2}, -\frac{1}{2}} \end{matrix}$
\\ \hline
brane scalar 
 & $\Phi_{({\bf 1}, {\bf 4})} : ({\bf 1}, {\bf 4})_{\frac{1}{2}} $ 
 & $\hat{\Phi} : ({\bf 1}, [{\bf 1,2}])_{\frac{1}{2}}$ 
\\ \hline
$\begin{matrix} {\rm symmetry ~of} \cr {\rm brane ~interactions} \end{matrix}$
 & $SU(3)_C \times SO(5) \times U(1)_X$ 
 & $SU(3)_C \times SO(4) \times U(1)_X$ 
\\ \hline
\end{tabular}
\label{Tab:matter}
\end{center}
}
\end{table}

The bulk action of each gauge field, $A_M^{SU(3)_C}$, $A_M^{SO(5)}$, or $A_M^{U(1)_X}$,
is given by 
\begin{align}
S_{\rm bulk}^{\rm gauge}&=
\int d^5x \sqrt{-\det G} \, \bigg[ -\tr \bigg( \frac{1}{4} F^{MN} F_{MN}
+ \frac{1}{2\xi}(f_{\rm gf})^2 + {\cal L}_{\rm gh} \bigg) \bigg],
\label{Action-bulk-gauge1}
\end{align}
where $\sqrt{-\det G}=1/k z^5$ and  $F_{MN} = \dd_M A_N - \dd_N A_M -i  g [A_M,A_N]$
with each 5D gauge coupling constant $g$.  
The gauge fixing $f_{\rm gf}$ and ghost terms ${\cal L}_{\rm gh}$ have been specified in ref.\  \cite{FHHOY2019a}.
Each fermion multiplet $\Psi (x,y)$ in the bulk has its own bulk-mass parameter $c$.\cite{Gherghetta2000}
The covariant derivative is given by
\begin{align}
&{\cal D}(c)= \gamma^A {e_A}^M
\bigg( D_M+\frac{1}{8}\omega_{MBC}[\gamma^B,\gamma^C]  \bigg) -c\sigma'(y) ~, \cr
\noalign{\kern 5pt}
&D_M =  \dd_M - ig_S A_M^{SU(3)} -i g_A A_M^{SO(5)}  -i g_B Q_X A_M ^{U(1)} ~. 
\label{covariantD1}
\end{align}
Here $\sigma' = d\sigma(y)/dy$ and $\sigma'(y) =k$ for $0< y < L$. 
$g_S$, $g_A$, $g_B$ are $SU(3)_C$, $SO(5)$, $U(1)_X$ gauge coupling constants.
The bulk part of the action for the fermion multiplets are given, with  $\overline{\Psi} = i \Psi^\dagger \gamma^0$, by
\begin{align}
S_{\rm bulk}^{\rm fermion} &=  \int d^5x\sqrt{-\det G} \, \bigg\{ 
\sum_J  \overline{\Psi}{}^J   {\cal D} (c_J) \Psi^J    \cr
\noalign{\kern 5pt}
&\hskip -0.5cm
-  \sum_\alpha \Big( m_{D_\alpha} \overline{\Psi}{}_{\bf (3,1)}^{+ \alpha} \Psi_{\bf (3,1)}^{- \alpha}  + {\rm H.c.} \Big)
-  \sum_\gamma \Big( m_{V_\gamma} \overline{\Psi}{}_{\bf (1,5)}^{+ \gamma} \Psi_{\bf (1,5)}^{- \gamma}  + {\rm H.c.} \Big)
 \bigg\} ,
\label{fermionAction1}
\end{align} 
where the sum $\sum_J$ extends over $ \Psi^J = \Psi_{\bf (3,4)}^\alpha$, $ \Psi_{\bf (1,4)}^\alpha$,
$ \Psi_{\bf (3,1)}^{\pm \alpha}$, $ \Psi_F^\beta$ and $ \Psi_{\bf (1,5)}^{\pm \gamma}$. 

The action for the brane scalar field $\Phi_{({\bf 1,4})}  (x)$  is given by
\begin{align}
S_{\rm brane}^{\Phi} & = 
\int d^5x\sqrt{-\det G} \,  \delta(y) \cr
\noalign{\kern 5pt}
&
\times \Big\{ 
-(D_\mu\Phi_{({\bf 1,4})})^{\dag}D^\mu\Phi_{({\bf 1,4})}
-\lambda_{\Phi_{({\bf 1,4})}}
\big(\Phi_{({\bf 1,4})}^\dag\Phi_{({\bf 1,4})} - |w|^2  \big)^2 \Big\} ,
\label{Action-brane-scalar1}
\end{align}
where $D_\mu = \dd_\mu - ig_A   A_{\mu}^{SO(5)}   - i \onehalf g_B   A_\mu^{U(1)}$.
The action for the gauge-singlet brane fermion $\chi^\alpha (x)$ is
\begin{align}
S_{\rm brane}^\chi = &
\int d^5x\sqrt{-\det G} \, \delta(y) \bigg\{  
\frac{1}{2}\overline{\chi}^\alpha \gamma^\mu\partial_\mu \chi^\alpha
 - \frac{1}{2} M^{\alpha \beta}  \overline{\chi}^\alpha \chi^{\beta} \bigg\} ~.
\label{brane-chi}
\end{align}
$\chi^\alpha (x)$ satisfies the Majorana condition
$\chi^c=\chi$;
\begin{align}
\chi = \begin{pmatrix} \xi \cr \eta \end{pmatrix} , ~~
\chi^c = \begin{pmatrix} + \eta^c \cr - \xi^c \end{pmatrix} 
=e^{i\delta_C} \begin{pmatrix} + \sigma^2 \eta^* \cr - \sigma^2 \xi^* \end{pmatrix} .
\label{Majorana1}
\end{align}

On the UV brane there are
$SU(3)_C\times SO(5) \times U(1)_X$-invariant
brane interactions among the bulk fermion,   brane fermion,
and brane scalar fields.  Relevant parts of the brane interactions  are given by
\begin{align}
 S_{\rm brane}^{\rm int} & =
- \int d^5x\sqrt{-\det G} \, \delta(y)  \times \cr
\noalign{\kern 5pt}
&
\Big\{  \kappa^{\alpha\beta} \, \overline{\Psi}{}_{({\bf 3,4})}^{\alpha} \Phi_{({\bf 1,4})} \cdot \Psi_{({\bf 3,1})}^{+\beta}  
+  \widetilde{\kappa}^{\alpha \beta} \, \overline{\chi}^\beta 
\widetilde{\Phi}_{({\bf 1,4})}^\dag \Psi_{({\bf 1,4})}^{\alpha}   + {\rm H.c.} \Big\} 
\label{brane-int1}
\end{align}
where $\kappa$'s and $\widetilde \kappa$'s are coupling constants and
\begin{align}
\Phi_{({\bf 1,4})} = \begin{pmatrix} \Phi_{[{\bf 2,1}]} \cr \Phi_{[{\bf 1,2}]} \end{pmatrix} , ~~
\widetilde{\Phi}_{({\bf 1,4})}= \begin{pmatrix}
i\sigma^2\Phi_{[{\bf 2,1}]}^* \cr \noalign{\kern 5pt}
-i\sigma^2\Phi_{[{\bf 1,2}]}^*  \end{pmatrix} .
\label{branescalar2}
\end{align}

When $\la  \Phi_{({\bf 1,4}) }\ra = (0,0,0,w)^t$,  (\ref{brane-int1}) generates additional mass terms
\begin{align}
 \int d^5x \sqrt{-\det G} \, \delta(y) \,  
\Big\{ 2 \mu^{\alpha \beta} \bar{d}_R^{\prime \alpha} D_L^{+\beta} + {\rm H.c.} \Big\} ~, ~~
\mu^{\alpha \beta}  = \frac{\kappa^{\alpha \beta} w}{\sqrt{2}} ~, 
\label{branemass1}
\end{align}
in the down-type quark sector  and
\begin{align}
- \int d^5x \sqrt{-\det G} \, \delta(y) \,     \frac{m_B^{\alpha\beta}}{\sqrt{k}} \, 
( \bar \chi^\beta \nu_R^{\prime \alpha} + \bar \nu_R^{\prime \alpha} \chi^\beta) ~, ~~
m_B^{\alpha\beta} = \widetilde \kappa^{\alpha \beta} w \sqrt{k} 
\label{branemass2}
\end{align}
in the neutrino sector.
With the Majorana masses in (\ref{brane-chi}), the mass term (\ref{branemass2})
induces inverse seesaw mechanism in the neutrino sector.\cite{HosotaniYamatsu2017}
Further $\la  \Phi_{({\bf 1,4}) }\ra \not= 0$ breaks $SO(4) \times U(1)_X$ down to $SU(2)_L \times U(1)_Y$.
We assume that $w \gg m_\KK$.
The 4D $SU(2)_L$ gauge coupling is given by $g_w = g_A/\sqrt{L}$.
The 5D gauge coupling $g_Y^{\rm 5D}$ of $U(1)_{Y}$  and the 4D bare Weinberg angle at the tree level,
$\theta_W^0$, are given by
\begin{align}
&g_Y^{\rm 5D} =\frac{g_Ag_B}{\sqrt{g_A^2+g_B^2}}  ~, \cr
\noalign{\kern 5pt}
&\sin \theta_W^0 = \frac{s_\phi}{\sqrt{\smash[b]{1 + s_\phi^2}}} ~, ~~
s_\phi  = \frac{g_B}{\sqrt{g_A^2+g_B^2}} ~.
\label{gY-sW}
\end{align}
The bare Weinberg angle $\theta_W^0$ 
with a given $\theta_H$  is determined  to fit the LEP1 data for $e^+ e^- \go \mu^+ \mu^-$ 
at $\sqrt{s} = m_Z$.\cite{LEPsummary}
Approximately $\sin^2 \theta_W^0 \simeq 0.1140 + 0.1186 \cos \theta_H   - 0.0014 \cos 2 \theta_H$.
Evaluated gauge couplings turn out very close to those in the SM
with $\sin^2 \theta_W = 0.2312$.\cite{FHHOY2019b}

The 4D Higgs boson $\Phi_H (x)$ is contained in the $SO(5)/SO(4)$ part of $A_y^{SO(5)}$.
 In the $z$ coordinate $A_z = (kz)^{-1} A_y$ ($1 \le z \le z_L$), and 
\begin{align}
A_z^{(j5)} (x, z) &= \frac{1}{\sqrt{k}} \, \phi_j (x) u_H (z) + \cdots , \cr
\noalign{\kern 5pt}
u_H (z) &= \sqrt{ \frac{2}{z_L^2 -1} } \, z ~,  \cr
\noalign{\kern 5pt}
\Phi_H (x) &= \frac{1}{\sqrt{2}} \begin{pmatrix} \phi_2 + i \phi_1 \cr \phi_4 - i\phi_3 \end{pmatrix} .
\label{4dHiggs}
\end{align}
At the quantum level $\Phi_H$ develops a nonvanishing expectation value.  Without loss of generality
we assume $\la \phi_1 \ra , \la \phi_2 \ra , \la \phi_3 \ra  =0$ and  $\la \phi_4 \ra \not= 0$, 
which is related to the Aharonov-Bohm (AB) phase $\theta_H$ in the fifth dimension.  Eigenvalues of 
\begin{align}
\hat W &= P \exp \bigg\{ i g_A \int_{-L}^L dy \, A_y \bigg\}  \cdot P_1 P_0 
\label{ABphase1}
\end{align}
are gauge invariant.  For $A_y = (2k)^{-1/2} \phi_4 (x)v_H (y) T^{(45)}$,  where
$v_H (y) = k e^{ky} u_H (z)$ for $0 \le y \le L$ and $v_H (-y) =  v_H (y) =  v_H (y + 2L)$,
one finds
\begin{align}
&\hat W = \exp \Big\{ i f_H^{-1} \phi_4 (x) \cdot 2 T^{(45)} \Big\} ~, \cr
\noalign{\kern 5pt}
&f_H = \frac{2}{g_A} \sqrt{ \frac{k}{z_L^2 -1}} = \frac{2}{g_w} \sqrt{ \frac{k}{L(z_L^2 -1)}} ~, \cr
\noalign{\kern 5pt}
&\theta_H = \frac{\la \phi_4 \ra}{f_H} ~.
\label{ABphase2}
\end{align}
Note
\begin{align}
&A_z^{(45)} (x, z) = \frac{1}{\sqrt{k}} \big\{ \theta_H f_H + H(x) \big\} \, u_H(z) + \cdots 
\label{ABphase3}
\end{align}
where $H(x)$ is the  neutral Higgs boson field.
There is a large gauge transformation which shifts $\theta_H$ by $2 \pi$, preserving the boundary 
conditions.  Physics is invariant under $\theta_H \go \theta_H + 2\pi$.
We shall evaluate the effective potential $V_\eff (\theta_H)$ in the next section.

\section{Effective potential}

The effective potential $V_\eff (\theta_H)$ at the one-loop level is evaluated 
from the mass spectra of all fields which depend on $\theta_H$.
After the Wick rotation into the Euclidean signature it is expressed as
\begin{align}
V_\eff(\theta_H) &=
\sum  \pm \frac{1}{2} \int \frac{d^4 p_E}{(2\pi)^4} \sum_n 
\ln \big\{ p_E^2 + m_n (\theta_H)^2 \big\} , 
\label{effVgeneral1}
\end{align}
where the sign $+$ ($-$) corresponds to bosons (fermions).
When the Kaluza-Klein (KK) spectrum $\{ m_n (\theta_H) \}$ is determined by 
zeros of a function $\rho (z; \theta_H)$, namely by
\begin{align}
\rho (m_n; \theta_H) = 0  ~~~ (n=1,2,3, \cdots) ,
\label{spectrum1}
\end{align}
then  $V_\eff (\theta_H)$ is given  \cite{Falkowski2007} by
\begin{align}
V_\eff (\theta_H) &= \sum \pm 
\frac{1}{(4\pi)^2}  \int_0^\infty dy \, y^3 \ln \rho (iy; \theta_H) ~.
\label{effVgeneral2}
\end{align}
The $\theta_H$-dependent part of $V_\eff^{1 \, {\rm loop}} (\theta_H)$ is finite,
independent of a cutoff and regularization method employed.

The spectrum-determining functions $\rho (z; \theta_H)$ for all fields in the model 
have been given in ref.\ \cite{FHHOY2019a}. 
They are summarized in Appendix A for convenience.
Relevant contributions come from $W$ and $Z$ gauge fields, top-bottom quark multiplets, 
and dark fermions in the spinor and vector representations.  
Contributions from light quarks and leptons are negligible.
To avoid unnecessary confusion in the following argument, we denote the effective potential 
as $V_\eff (\theta)$. Physical value $\theta_H$ corresponds to the global minimum of 
$V_\eff (\theta)$, namely $dV_\eff /d \theta  |_{\theta= \theta_H} = 0$.
One finds 
\begin{align}
V_\eff (\theta) &=
 2(3-\xi^2) A_W(\theta) + (3-\xi^2) A_Z(\theta)+ 3 \xi^2 A_S(\theta)  \cr
\noalign{\kern 5pt}
&  -12 A_{\rm top}(\theta)  -12 A_{\rm bottom}(\theta) 
- 12 n_F A_F(\theta) - 8 n_V A_V(\theta) ~, \cr
\noalign{\kern 5pt}
A_p(\theta) &\equiv \frac{(k z_L^{-1})^4}{(4\pi)^2}
\int_0^\infty dq \, q^3 \ln \bigg\{ 1 + \sum_{n=1}^2 Q_p^{(n)}(q)\cos(n \theta) \bigg\} ,
\label{effV1}
\end{align}
where $n_F$ and $n_V$ are the number of $\Psi_F$ and $\Psi_{(1,5)}^\pm$, and 
$\xi$ is a gauge parameter in the generalized $R_\xi$ gauge.
The integration variable has been changed from $y$ in (\ref{effVgeneral2}) to $q = k^{-1} z_L y$.
In the following we take $z_L=e^{k L}=10^{10}$ and $\xi=0$.
The contributions from $W$, $Z$ towers and Goldstone boson tower are given by 
\begin{align}
&Q_W^{(1)}(q) =  Q_Z^{(1)}(q) =   Q_S^{(1)}(q) = 0 ~, \cr
\noalign{\kern 5pt}
& Q_W^{(2)}(q) = - \frac{1}{- 4 i  z_Lq^{-1}  \hat{C}^\prime (q) \hat{S}(q) + 1} ~,  \cr
\noalign{\kern 5pt}
& Q_Z^{(2)}(q) = - \frac{ 1 + s_\phi^2 }{- 4 i  z_L q^{-1}  \hat{C}^\prime (q) \hat{S}(q) + 1 + s_\phi^2 } ~, \cr
\noalign{\kern 5pt}
& Q_S^{(2)}(q) = - \frac{1}{- 2i z_L q^{-1} \hat{C}(q) \hat{S}(q) + 1} ~.
\label{effVgauge}
\end{align}
$\hat C(q)$, $\hat S (q)$ etc.\ in the expressions above and $\hat C_L (q,c)$, 
$\hat S_L (q,c)$ etc.\ in the  expressions below are  given in Appendix B.  They are expressed
in terms of modified Bessel functions.

Top and bottom quark contributions are given by 
\begin{align}
& Q_{\rm top}^{(2)}(q) = Q_{\rm bottom}^{(2)}(q) = 0 ~, \cr
\noalign{\kern 5pt}
&Q_{\rm top}^{(1)}(q) =  - \frac{1}{2 \hat{S}_L(q; c_t) \hat{S}_R(q; c_t) +1} ~, \cr
\noalign{\kern 5pt}
& Q_{\rm bottom}^{(1)}(q) =   -\frac{\widetilde{S_LS_R}}
{2 \hat{S}_L(q; c_t) \hat{S}_R(q; c_t) \widetilde{S_LS_R}
  +2 | \mu |^2 \hat{C}_R(q; c_t) \hat{S}_R(q; c_t)\widetilde{C_LS_L} +1 } ~, \cr
\noalign{\kern 5pt}
&  \widetilde{S_LS_R} = 
\hat{S}_L(q; c_{D_b} + \tilde{m}_{D_b}) \hat{S}_R(q; c_{D_b}- \tilde{m}_{D_b})
+\hat{S}_L(q; c_{D_b} - \tilde{m}_{D_b}) \hat{S}_R(q; c_{D_b} + \tilde{m}_{D_b}) \cr
\noalign{\kern 5pt}
&\hskip .5cm
+\hat{C}_L(q; c_{D_b} + \tilde{m}_{D_b}) \hat{C}_R(q; c_{D_b} - \tilde{m}_{D_b})
+\hat{C}_L(q; c_{D_b} - \tilde{m}_{D_b}) \hat{C}_R(q; c_{D_b} + \tilde{m}_{D_b})  -2 ~,  \cr
\noalign{\kern 5pt}
&\widetilde{C_LS_L} =
\hat{C}_L(q; c_{D_b} + \tilde{m}_{D_b}) \hat{S}_L(q; c_{D_b} - \tilde{m}_{D_b})
+\hat{C}_L(q; c_{D_b} - \tilde{m}_{D_b}) \hat{S}_L(q; c_{D_b} + \tilde{m}_{D_b}) ~. 
\label{effVtop}
\end{align}
In the above expressions we have assumed that the brane interaction term (\ref{brane-int1}) is 
diagonal in generation space.  $c_{D_\alpha}$ is the bulk mass parameter of 
$\Psi_{(3,1)}^{\pm \alpha}$ and $\tilde{m}_{D_\alpha} = m_{D_\alpha} /k$.
Numerically the contribution of bottom quark is very small and may be ignored. 
There are two kinds of dark fermions ($\Psi_F^\beta$ and $\Psi_{(1,5)}^{\pm \gamma}$).
Their contributions are  given by  
\begin{align}
& Q_{F}^{(1)}(q) =  \frac{1}{2 \hat{S}_L(q; c_F) \hat{S}_R(q; c_F) +1} ~, \cr
\noalign{\kern 5pt}
& Q_{F}^{(2)}(q) =  0 ~, \cr
\noalign{\kern 5pt}
&Q_{V}^{(1)}(q) =  0 ~, \cr
\noalign{\kern 5pt}
&Q_{V}^{(2)}(q) =  - \frac{2}{\hat{B}_0(q; c_V,  \tilde{m}_V)} ~, \cr
\noalign{\kern 5pt}
&\hat{B}_0(q; c_V,  \tilde{m}_V) = \cr
\noalign{\kern 5pt}
&\quad
~~ \hat{C}_L(q; c_V +  \tilde{m}_V) \hat{C}_R(q; c_V -  \tilde{m}_V) 
  + \hat{C}_L(q; c_V -  \tilde{m}_V) \hat{C}_R(q; c_V +  \tilde{m}_V)  \cr
\noalign{\kern 5pt}
&\quad
+ \hat{S}_L(q; c_V +  \tilde{m}_V) \hat{S}_R(q; c_V -  \tilde{m}_V) 
+ \hat{S}_L(q; c_V -  \tilde{m}_V) \hat{S}_R(q; c_V +  \tilde{m}_V) ~.
\label{effVdark}
\end{align}
For the sake of simplicity we set degenerate bulk mass parameters $c_F$ for $\Psi_F^\beta$,
and degenerate masses $m_V = k \tilde m_V$ and bulk mass parameters $c_V$ for
$\Psi_{(1,5)}^{\pm \gamma}$.
We note that contributions from gauge bosons and $\Psi_{(1,5)}^{\pm \gamma}$ fields to $V_\eff (\theta)$
are periodic in $\theta$ with a period  $\pi$, whereas those from top-bottom quarks and $\Psi_F^\beta$ 
fields are periodic with a period $2 \pi$.

The parameters of the model are determined in the following steps.  
(i) We pick the value of $\theta_H$.  In other words we are going to adjust the parameters of the model 
such that  $V_\eff (\theta)$ has a global minimum at $\theta=\theta_H$.   (ii) We take $z_L = 10^{10}$.  
Then $k$ is determined for $m_Z$ to be reproduced, and
the KK mass scale $m_\KK = \pi k (z_L -1)^{-1}$ is fixed.  
(iii) The bulk mass parameters of $\Psi_{(3,4)}^\alpha$ and $\Psi_{(1,4)}^\alpha$ are 
fixed from the masses of up-type quarks and charged leptons.  In particular, $c_t$ is determined by $m_t$.  
(iv) The bulk mass parameters $c_{D_\alpha}$ of $\Psi_{3,1)}^{\pm \alpha}$ and brane interaction
coefficients $\mu^{\alpha \beta}$ are determined so as to reproduce the masses of down-type quarks 
and CKM matrix.  Similarly the Majorana mass terms $M^{\alpha\beta}$ and brane interctions 
$\tilde\kappa^{\alpha\beta}$ are determined so as to reproduce neutrino masses and 
PMNS matrix.  As remarked above, these parameters are numerically irrelevant for $V_\eff (\theta)$.
(v) At this stage there remain five parameters to be determined;  $(n_F, c_F)$ of  $\Psi_F^\beta$ 
and $(n_V, c_V, \tilde m_V)$ of $\Psi_{(1,5)}^{\pm \gamma}$.
There are two conditions to be satisfied;
\begin{align}
(a):&\quad  
\frac{dV_\eff}{d\theta} \bigg|_{\theta=\theta_H} = 0 ~, \cr
\noalign{\kern 5pt}
(b):&\quad  
m_H^2 = \frac{1}{f_H^2} \frac{d^2 V_\eff}{d \theta^2} \bigg|_{\theta=\theta_H} ~,
\label{HiggsMass1}
\end{align}
where $m_H = 125.1\,$GeV.
The second condition for the Higgs boson mass $m_H$ follows from the fact that 
the effective potential for the 4D Higgs field $H(x)$ is given by $V_\eff (\theta_H + f_H^{-1} H)$
as infered from (\ref{ABphase3}).
The  conditions  (\ref{HiggsMass1})  give two constraints to be satisfied among the five
parameters $(n_F, c_F, n_V, c_V, \tilde m_V)$.
We first fix, for instance,  $(n_F, n_V, c_V)$ and determine $(c_F, \tilde m_V)$ by (\ref{HiggsMass1}).

One may wonder whether the arbitrary choice of the parameters in the last step
diminishes prediction power  of the model.  Quite surprisingly  many of the physical quantities
do not depend on such details in the parameter choice, being determined solely by $\theta_H$.
There appears the $\theta_H$-universality which will be explained in the next section.

\begin{figure}[htb]
\begin{tabular}{cc}
\begin{minipage}{0.5\hsize}
\begin{center}
\includegraphics[width=75mm]{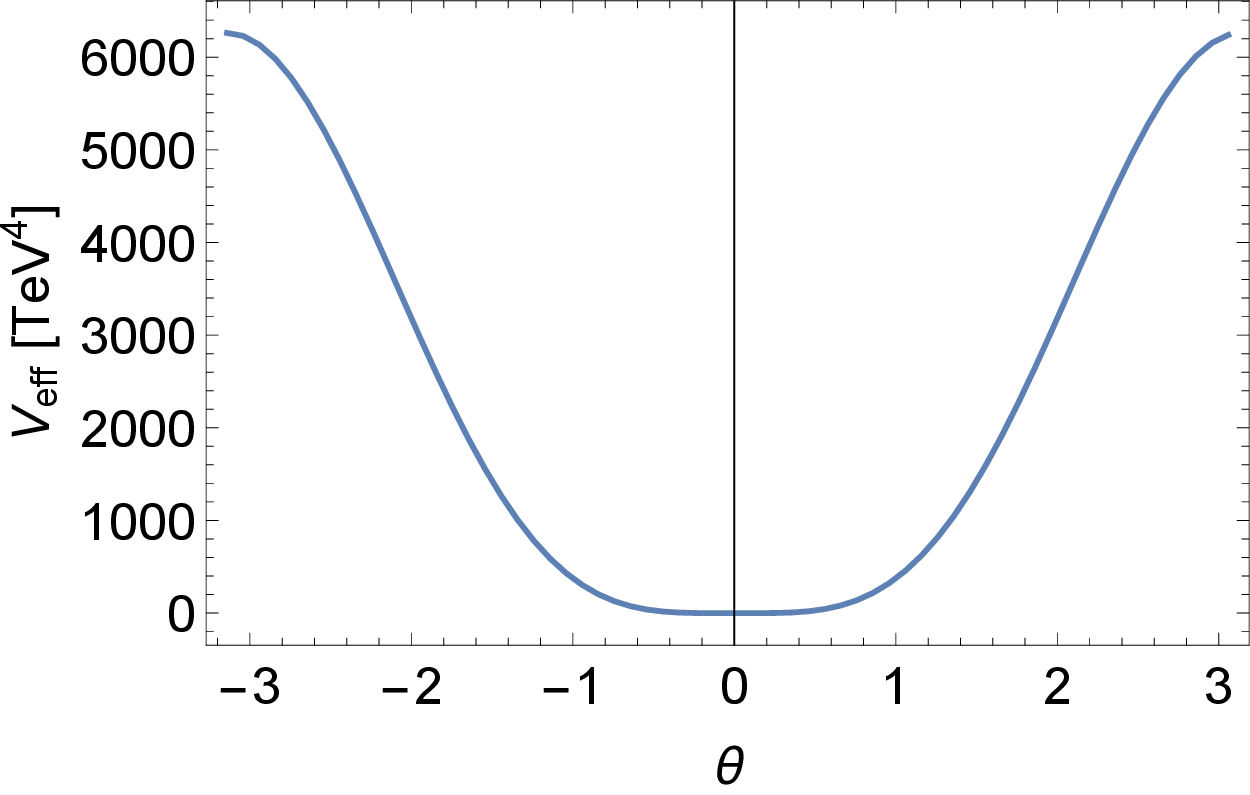}
\end{center}
\end{minipage}
\begin{minipage}{0.5\hsize}
\begin{center}
\includegraphics[width=75mm]{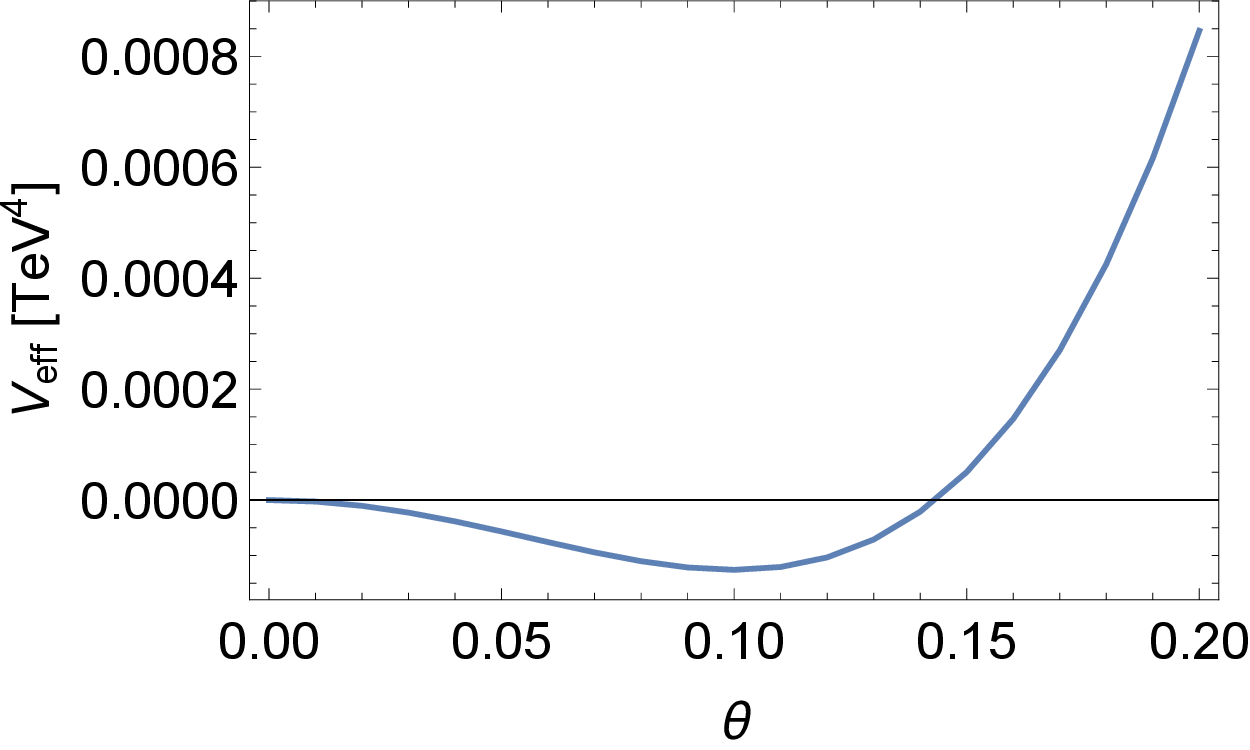}
\end{center}
\end{minipage}
\end{tabular}
\caption{The effective potential for $\theta_H=0.1$, $n_F=2$, $n_V= 2$ and $c_V=0$.
The global minimum is located at $\theta=\theta_H$.}
\label{fig:pot}
\end{figure}

\begin{figure}[tbh]
\centering
\includegraphics[width=100mm]{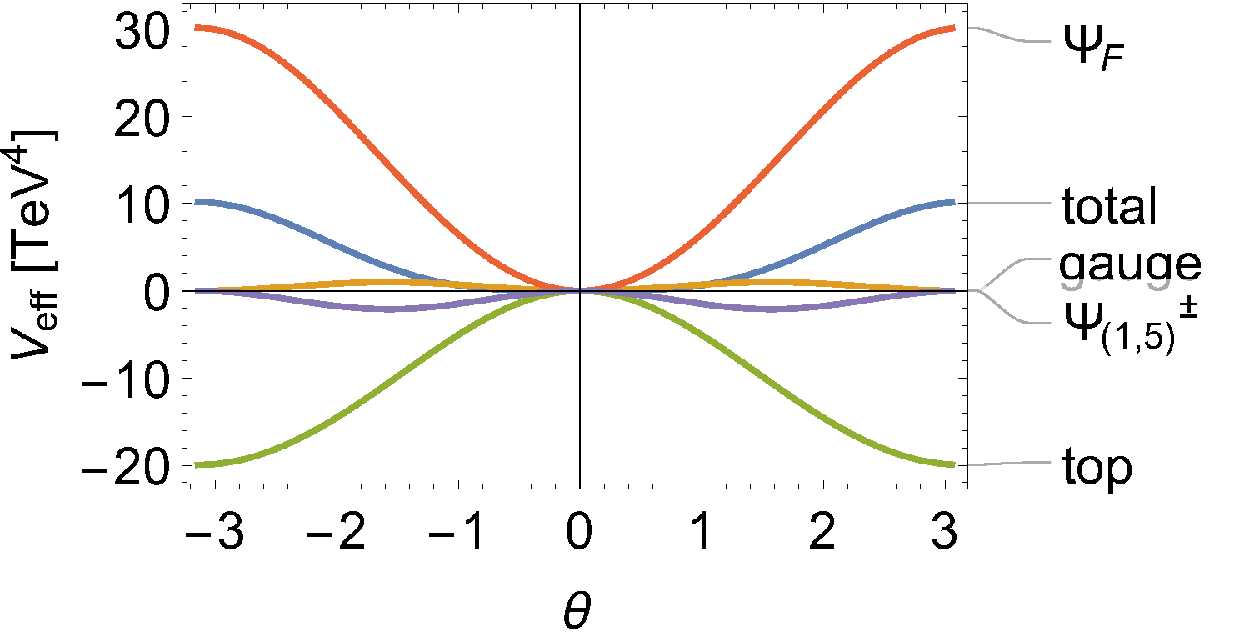}
\caption{Contributions of relevant fields to the effective potential
for $\theta_H=0.1$, $n_F=2$, $n_V= 2$ and $c_V=0$ are displayed.
}   
\label{fig:pot2}
\end{figure}

We give some examples.  The parameters fixed in the steps (i) to (iv) above are tabulated
in Table \ref{Tab:parameter1}.
In Fig.\ \ref{fig:pot} the effective potential for $\theta_H=0.1$, $n_F=n_V= 2$ and $c_V=0$ 
is displayed.  $c_F = 0.319$ and $\tilde{m}_V = 0.0806$ are chosen to satisfy (\ref{HiggsMass1}) . 
One observes that the electroweak symmetry is dynamically broken.  
In Fig.\ \ref{fig:pot2} contributions of relevant fields to the effective potential $V_\eff (\theta)$ 
are displayed.  
There is a lower bound for $\theta_H$ in order to reproduce the top quark mass.
$\theta_H \ge \theta_{c1}$ where $\theta_{c1} \sim 0.015$ for $z_L = 10^{10}$.  
Similarly there is a constraint for the warp factor.
For $\theta_H = 0.1$, $n_F = n_V =2$, the top quark mass is reproduced only if 
$z_L \ge z_{L1} \sim 10^{8.1}$ and dynamical electroweak symmetry breaking is achieved
only if $z_L \le z_{L2} \sim 10^{15.5}$.

\begin{table}[bht]
{
\renewcommand{\arraystretch}{1.2}
\begin{center}
\caption{Parameters determined for $z_L=10^{10}$ and $\theta_H=0.05, 0.1, 0.15$.
We have set $\mu^{\alpha\beta} = \mu_\alpha \delta_{\alpha\beta}$, and 
have taken $c_{D_b} = 1.04$.   $\mu_b$ is  determined  so as to reproduce $m_b$. }
\vskip 10pt
\begin{tabular}{|c||c|c|c|c|c|}
\hline
$\theta_H$ & $k$ [$10^{13}$ GeV] &$m_\KK$ [TeV] & $c_t$ & $c_{D_b}$ &$\mu_{b}$  \\
\hline
0.05 & 7.68 & 24.1  & -0.226 & 1.04 & 0.106   \\
\hline
0.10 & 3.84 & 12.1  & -0.227 & 1.04 & 0.104   \\
\hline
0.15 & 2.57 & 8.07  & -0.230 & 1.04 & 0.0990 \\
\hline
\end{tabular}
\label{Tab:parameter1}
\end{center}
}
\end{table}

\begin{figure}[tbh]
\centering
\includegraphics[width=100mm]{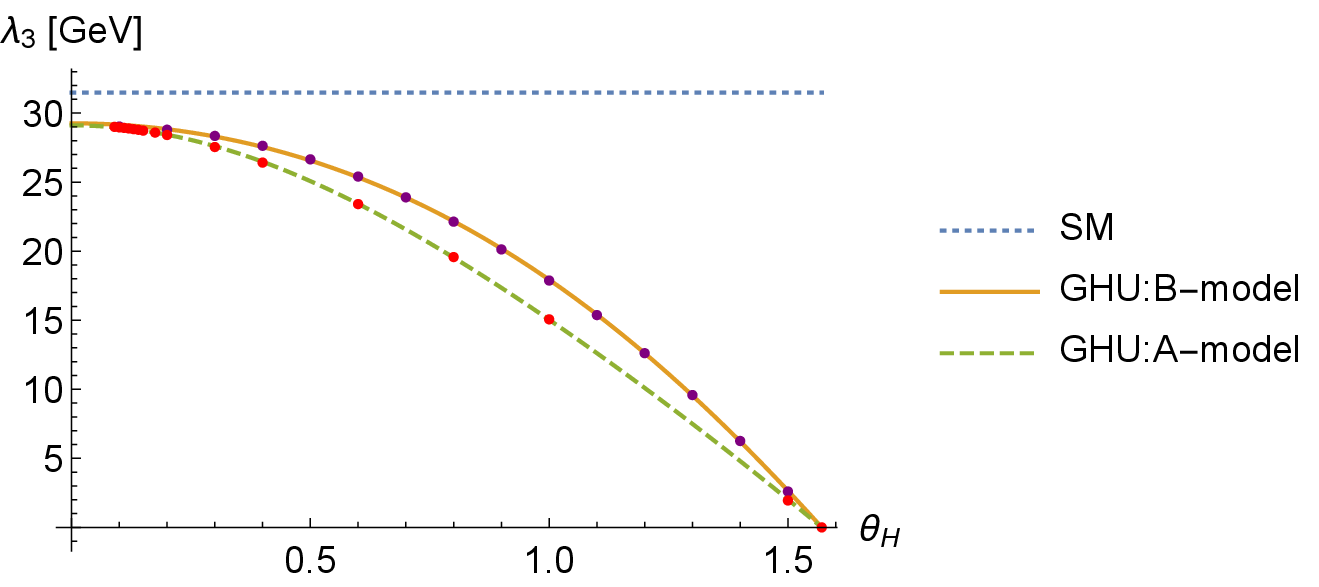}
\caption{
 The cubic coupling $\lambda_3$ of the Higgs boson. 
 The fitting curves are given by 
 $\lambda_3/{\rm GeV} = 39.6 \cos \theta_H  - 5.21 (1 + \cos 2\theta_H ) - 0.00911 \cos 3\theta_H $ for the B-model 
 and  $\lambda_3/{\rm GeV} = 32.4 \cos \theta_H  - 2.26 (1 + \cos 2\theta_H)  - 1.1 \cos 3\theta_H $ for the A-model.
 The SM value is $\lambda_{3, SM} = 31.5 \, {\rm GeV}$.
}   
\label{fig:self3}
\end{figure}

\begin{figure}[tbh]
\centering
\includegraphics[width=100mm]{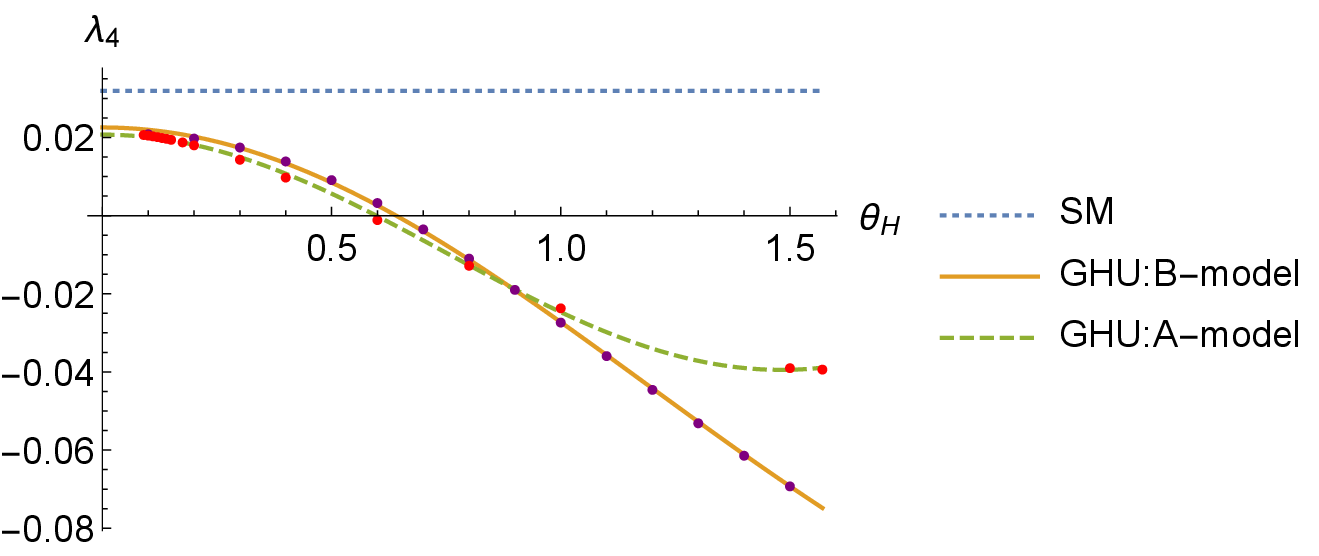}
\caption{
 The quartic coupling $\lambda_4$ of the Higgs boson.
 The fitting curves are given by $\lambda_4 = -0.0695 + 0.0852 \cos \theta_H  + 0.00725 \cos 2\theta_H $ for the B-model 
 and   $\lambda_4 = -0.00264 - 0.0129 \cos \theta_H  + 0.0363 \cos 2\theta_H$ for the A-model.
 The SM value is $\lambda_{4, SM}= 0.0320$. 
}
\label{fig:self4}
\end{figure}

The effective potential $V_\eff (\theta)$ has more information.
By expanding $V_\eff(\theta_H + H/f_H)$, one finds Higgs self-couplings $\lambda_n H^n$. 
The $n$-th self-coupling $\lambda_n$ is given by 
\begin{eqnarray}
\lambda_n \equiv \frac{1}{n! f_H^n} \frac{d^n V_\eff}{d \theta^n} \bigg|_{\theta=\theta_H} .
\label{selfcouplings1}
\end{eqnarray}
The couplings $\lambda_3$ and $\lambda_4$ are plotted in Fig.\ \ref{fig:self3} and Fig.\ \ref{fig:self4}
as functions  of $\theta_H$ for $c_V=0.2, n_F=n_V = 2$.
The fitting curves are given by
\begin{align}
{\rm B~model:}~
&\lambda_3 /{\rm GeV} =  39.6 \cos \theta_H - 5.21 (1 + \cos 2\theta_H)  - 0.00911 \cos 3\theta_H  ~, \cr
\noalign{\kern 5pt}
&\lambda_4 =  -0.0695 + 0.0852 \cos \theta_H  + 0.00725 \cos 2\theta_H ~.
\label{Hselfcoupling-B}
\end{align}
In Figs.\ \ref{fig:self3} and  \ref{fig:self4}, $\lambda_3$ and $\lambda_4$ in the A model are also
plotted, for which the fitting curves are given by
\begin{align}
{\rm A~model:}~
&\lambda_3 /{\rm GeV} =  32.4 \cos \theta_H  - 2.26 (1 + \cos 2\theta_H)  - 1.1 \cos 3\theta_H ~, \cr
\noalign{\kern 5pt}
&\lambda_4 =  -0.00264 - 0.0129 \cos \theta_H  + 0.0363 \cos 2\theta_H ~.
\label{Hselfcoupling-A}
\end{align}
Note that $\lambda_3$ vanishes at $\theta_H = \onehalf \pi$ as a consequence of 
the $H$ parity in GHU models.\cite{Hparity}
For $\theta_H \gtrsim 0.6$,  $\lambda_4$ becomes negative, which, however, does not mean 
the instability.  The $\theta$-dependent part of $V_\eff (\theta)$ is finite, bounded from below.
In gauge-Higgs unification there does not arise the vacuum instability problem which afflicts
most of 4D field theories.
From the experimental constraints from the LEP1, LEP2 data and from the LHC data for the
nonobservation of $Z'$ events it is infered that $\theta_H \lesssim 0.11$.
For $\theta_H \sim 0.1$ ($0.15)$, $\lambda_3$ and $\lambda_4$ are smaller than those in the
SM by 7.7\% (8.1\%)  and 30\% (32\%),  respectively.
As explained above, there is a lower bound for $\theta_H$ in GHU, namely $\theta_H > \theta_{c1}$, 
and there does not exist $\theta_H \go 0 $ limit.  It is not surprising that $\lambda_3$ and $\lambda_4$ deviate
from the values in the SM even for small $\theta_H$.
The couplings $\lambda_n$ ($n \ge 5$) are generated at the one loop level both in GHU and in the SM,
which turn out finite.  The Higgs couplings to quarks, leptons, and $W$, $Z$ bosons in GHU for small $\theta_H$ 
are very close to those in the SM.  There are additional contributions coming from KK modes in GHU.
It may be interesting as well to measure these couplings $\lambda_n$ ($n \ge 5$) in future experiments to test
predictions from GHU and the SM.


\section{$\theta_H$ universality}

As remarked in the previous section there remains the arbitrariness in the choice of
the parameters in the model.  
Among the five parameters $(n_F, c_F, n_V, c_V, \tilde m_V)$ there are only two conditions 
in  (\ref{HiggsMass1}) to be obeyed.  In the examples given in the previous section
we fisrt fixed $(n_F, n_V, c_V)$ and determined $(c_F, \tilde m_V)$ by (\ref{HiggsMass1}).
The Higgs cubic and quartic couplings, $\lambda_3$ and $\lambda_4$,  are evaluated with this choice.
One might wonder how $\lambda_3$ and $\lambda_4$ depend on the choice of the parameters
$(n_F, n_V, c_V)$.

In this section we shall show that $\lambda_3$ and $\lambda_4$ are determined, to high accuracy,  
as functions of $\theta_H$ only, but do not depend on the details of the parameter choice.
It has been known in GHU that the 3-point Higgs couplings
to $W$, $Z$, quarks and leptons also have the same property.\cite{HK2009}
These physical quantities are determined by $\theta_H$ to high accuracy.   
It may be  called as the $\theta_H$ universality.
The $\theta_H$ universality leads to profound power for predictions.  
Once the value of $\theta_H$ is determined by one of the physical quantities, then
the values of   other physical quantities are predicted.

In Table \ref{tab:universality} evaluated values of $(\lambda_3, \lambda_4)$ for $\theta_H = 0.1$
are shown with various choices of $(n_F, n_V, c_V)$.  
Although the values of determined $c_F$ and $\tilde m_V$ depend on the choice of $(n_F, n_V, c_V)$,
the evaluated values of $\lambda_3$ and $\lambda_4$ are universal to high accuracy.
$\lambda_3$ and $\lambda_4$ are determined as functions of $\theta_H$ only.

\begin{table}[tb]
\renewcommand{\arraystretch}{1.1}
\centering
\caption{
$\theta_H$ universality in $\lambda_3$ and $\lambda_4$ for $\theta_H=0.1$ and $z_L = 10^{10}$.  
With given $(n_F, n_V, c_V)$, $c_F$ and $ \tilde m_V$ are determined to satisfy 
the condition (\ref{HiggsMass1}), and $\lambda_3$ and $\lambda_4$ are evaluated by (\ref{selfcouplings1}).
}
\vskip 10pt
\begin{tabular}{|c|c|c||c|c||c|c|}
\hline
$n_F$ & $n_V$ & $c_V$ & $c_F$ & $\tilde m_V$ & $\lambda_3$(GeV) & $\lambda_4$ \\
\hline
2  & 2 & 0. & 0.319 & 0.0806 & 29.03 & 0.02083 \\
\hline
2  & 2 & 0.2 & 0.319 & 0.0777 & 29.03 & 0.02083 \\
\hline
2 & 2 & 0.5 & 0.322 & -0.0371  & 29.02 & 0.02078 \\
\hline
4 & 2 & 0. & 0.425 & 0.0794  & 29.02 & 0.02082 \\
\hline
4  & 2 & 0.2 & 0.425 & 0.0765 & 29.02 & 0.02082 \\
\hline
4  & 2 & 0.5   & 0.426 & -0.0350  & 29.01 & 0.02076 \\
\hline
2 & 4 & 0. & 0.318 & 0.0964  & 29.03 & 0.02084 \\
\hline
2  & 4 & 0.2 & 0.318 & 0.0937 & 29.03 & 0.02084 \\
\hline
2  & 4 & 0.5   & 0.319 & 0.0615  & 29.03 & 0.02083 \\
\hline
\end{tabular}
\label{tab:universality}
\end{table}

There is a reason for the $\theta_H$ universality.  
We first examine  global behavior of $V_\eff (\theta)$ with a given $\theta_H$.
Notice that the function $A_p (\theta)$ in (\ref{effV1}) is expanded as 
\begin{align}
 A_p(\theta) &=  \frac{(k z_L^{-1})^4}{(4\pi)^2} 
 \sum_{\ell =1}^\infty \sum_{n=1}^2 \alpha_p^{(n, \ell)} \cos^ \ell  n \theta ~,  \cr
 \noalign{\kern 5pt}
\alpha_p^{(n, \ell)}  &\equiv \frac{(-1)^{\ell +1}}{\ell}
	\int_0^\infty dq q^3   \left( Q_p^{(n)}(q)\right)^\ell ~.
\label{effV2}
\end{align}
As either $Q_p^{(1)}(q) $ or $Q_p^{(2)}(q)$ with given $p$ vanishes in (\ref{effV1}), 
$\alpha_p^{(1, \ell)} $ or $\alpha_p^{(2, \ell)} = 0$ for each $p$ in (\ref{effV2}).

To understand qualitative behavior of $V_\eff (\theta)$,  let us approximate 
$A_p (\theta)$ in (\ref{effV1}) by
\begin{align}
 \frac{(4\pi)^2}{(k z_L^{-1})^4} \, A_p(\theta) = \begin{cases} 
\alpha_p^{(2, 1)} \cos2 \theta & {\rm for~}  p= W, Z, S, V, \cr
\noalign{\kern 5pt}
\alpha_p^{(1, 1)} \cos\theta+\alpha_p^{(1, 2)} \cos^2\theta  & {\rm for~} p= {\rm top}, F. 
\end{cases}
\label{effV3}
\end{align}
Note that $|\alpha_p^{(n, 2)} / \alpha_p^{(n, 1)} | < 0.05$.
As the contributions from $p= {\rm top}, F$ are one order of magnitude larger than
those from $p= W, Z, S, V$,  the $\cos^2\theta$ terms have been retained for top and $F$.
$V_\eff (\theta)$ in this approximation, denoted as $V_\app (\theta)$,  is given by 
\begin{align}
&
V_\app(\theta) 
=\frac{(k z_L^{-1})^4}{(4\pi)^2} \Big(-  B_1 \cos\theta + B_2 \cos2\theta  \Big)~,  \cr
\noalign{\kern 5pt}
&B_1 = 12 \alpha_{\rm top}^{(1, 1)} + 12 n_F \alpha_F^{(1, 1)}  ~, \cr
\noalign{\kern 5pt}
&B_2 = \alpha_{\rm gauge} - 8 n_V \alpha_V^{(2, 1)} -6 \alpha_{\rm top}^{(1, 2)} 
- 6 n_F \alpha_F^{(1, 2)}~, \cr
\noalign{\kern 5pt}
&\alpha_{\rm gauge} = 2(3-\xi^2) \alpha_W^{(2, 1)}+ (3-\xi^2) \alpha_Z^{(2, 1)}
+ 3 \xi^2 \alpha_S^{(2, 1)} ~. 
\label{effV4a}
\end{align}
The condition $(a)$ in (\ref{HiggsMass1}) leads to $B_1 = 4 B_2 \cos\theta_H$.
Then the condition $(b)$  in (\ref{HiggsMass1}) implies that 
\begin{align}
B_2 \sim \frac{16 \pi^2}{g_w^2 (kL)^2} \Big( \frac{m_H}{m_W} \Big)^2
\label{effV4b}
\end{align}
where the relations $m_\KK \sim \pi k z_L^{-1}$, $m_W \sim (k/L)^{1/2} z_L^{-1} \sin \theta_H$,
$f_H \sim 2 m_W/ g_w \sin \theta_H$ have been made use of.
It follows that 
\begin{align}
&V_\app (\theta) =  V_0 \, u(\theta) ~, \cr
\noalign{\kern 5pt}
& u(\theta) = - 4 \cos \theta_H   \cos \theta + \cos 2 \theta    ~, \cr
\noalign{\kern 5pt}
&V_0 = \frac{m_W^2 m_H^2}{g_w^2 \sin^4 \theta_H} ~.
\label{effVapprox}
\end{align}
The cubic and quartic Higgs self-couplings are given by
\begin{align}
\lambda_3^\app &\sim \frac{g_w m_H^2}{4 m_W} \, \cos \theta_H ~, \cr
\noalign{\kern 5pt}
\lambda_4^\app &\sim \frac{g_w^2 m_H^2}{96 m_W^2} \, ( 7 \cos^2 \theta_H - 4) ~.
\label{Hselfcoupling-approx}
\end{align}

The approximate formulas (\ref{effVapprox}) and (\ref{Hselfcoupling-approx}) represent
qualitative behavior of the effective potential $V_\eff (\theta)$, but exhibit slight deviation
from the values in Table \ref{tab:universality} and the fitting curves (\ref{Hselfcoupling-B}) and
(\ref{Hselfcoupling-A}).  We first note that the form of $V_\app (\theta)$ is fixed,
once one makes an Ansatz that $V_\eff (\theta)$ is expressed in terms of two functions $\cos\theta$ and
$\cos 2 \theta$.  The relevant quantities are $B_1$ and $B_2$, but not detailed values of the 
parameters in the models considered.  In the A- or B-model the same universality relations 
(\ref{Hselfcoupling-approx}) result in this approximation.
It is easy to confirm that the formulas (\ref{Hselfcoupling-approx}) reproduce the SM values
at $\theta_H = 0$; 
\begin{align}
&\lambda_3^\app \big|_{\theta_H = 0} = \lambda_{3, SM} ~, ~~
\lambda_4^\app  \big|_{\theta_H = 0} = \lambda_{4, SM} ~.
\label{Hselfcoupling-SMlimit}
\end{align}
We also note that $u(\pi) - u(0) = 8 \cos\theta_H$ and $u(\theta_H) - u(0) = -2 (1-\cos\theta_H)^2$.
For small $\theta_H$, $u(\pi) - u(0) \sim  8$ and $u(\theta_H) - u(0) \sim - \onehalf \theta_H^4$,
which explains the behavior of $V_\eff (\theta)$ for $\theta_H = 0.1$ seen in fig.\ \ref{fig:pot}.

To understand the $\theta_H$ universality demonstrated in the previous section, 
refinement of the arguments is necessary.   The universality was first found in the A-model 
of $SO(5) \times U(1)$ gauge-Higgs unification.\cite{FHHOS2013} 
The mechanism for yielding the $\theta_H$ universality has been explained in ref.\ \cite{GHbook1}.
We generalize the argument for the current B-model.
The important observation is that $\lambda_3$ and $\lambda_4$ are determined by the 
local behavior of the effective potential $V_\eff (\theta_H)$ in the vicinity of the global
minimum at $\theta=\theta_H$, and the universality reflects the local, but not global behavior of 
$V_\eff (\theta_H)$.

The effective potential $V_\eff (\theta)$, (\ref{effV1}), is decomposed into three parts;
\begin{align}
&V_\eff (\theta) = \frac{(k z_L^{-1})^4}{(4\pi)^2} \Big\{
h_0  (\theta) + n_F h_F (\theta; c_F, z_L) 
+ n_V h_V (\theta;  c_V, \tilde m_V, z_L)  \Big\}  
\label{universality1}
\end{align}
where $h_0(\theta)$ represents the contributions from gauge and top quark fields.
With   $\theta_H$, $z_L$ and $\xi$ specified, $k$ is determined from $m_Z$ and
$c_t$ is subsequently determined by $m_t$ so that $h_0 (\theta)$ is fixed.
All other parameters associated with quarks and leptons are  irrelevant for $V_\eff (\theta)$.
There remain five parameters $(n_F, c_F, n_V, c_V, \tilde m_V)$ to be specified in (\ref{universality1}).
They must be adjusted such that the two conditions in (\ref{HiggsMass1}) are satisfied.
The important feature in the RS space with $z_L \gg 1$ is that  the $\theta$ dependence of
$h_F (\theta; c_F, z_L) $ and $h_V (\theta; c_V, \tilde m_V, z_L )$ factorizes near $ \theta = \theta_H$
\begin{align}
&h_F (\theta; c_F, z_L)  \simeq \alpha_F (c_F, z_L) \tilde h_F (\theta) ~, \cr
&h_V (\theta; c_V, \tilde m_V, z_L)  \simeq \alpha_V (c_V, \tilde m_V, z_L) \tilde h_V (\theta) 
\label{universality2}
\end{align}
to very high accuracy.  This can be confirmed numerically from the formula for $A_p(\theta)$  in (\ref{effV1}).
The relation (\ref{universality2}) implies, for instance, that the ratio
$h_F (\theta; c_F^{(1)}, z_L) / h_F (\theta; c_F^{(2)}, z_L)$
is $\theta$-independent  near $\theta_H$.
For $\theta_H=0.1$, $z_L = 10^{10}$, and $(c_F^{(1)}, c_F^{(2)}) = (0.3, 0.4)$,
the ratio varies from 1.7916 to 1.7896  in the range $0.09 \le \theta \le 0.11$.
The variation is only 0.1\%.
We stress that this factorization formulas are valid only locally, namely near $ \theta = \theta_H$,
and $\tilde h_F (\theta)$ and $\tilde h_V (\theta)$ depend on $\theta_H$.
The $z_L$ dependence of $h_F (\theta ; c_F, z_L)$ and $h_V (\theta ; c_V, \tilde m_V, z_L)$ is also
tiny in the range $10^8 \lesssim z_L \lesssim 10^{15}$.

Let us pick a set of values $(n_F, n_V, c_V)$ and determine  $(c_F, \tilde m_V) $ by (\ref{HiggsMass1}).  
Making use of (\ref{universality2}), 
one finds
\begin{align}
&\bigg[ \frac{d h_0}{d \theta} + n_F \alpha_F (c_F, z_L)  \frac{d \tilde h_F}{d\theta} 
+ n_V  \alpha_V (c_V, \tilde m_V, z_L)  
\frac{d \tilde h_V}{d \theta}\bigg]_{\theta = \theta_H} = 0 ~, \cr
\noalign{\kern 10pt}
&\bigg[ \frac{d^2 h_0}{d \theta^2} + n_F \alpha_F (c_F, z_L)  \frac{d^2 \tilde h_F}{d\theta^2} 
+ n_V  \alpha_V (c_V, \tilde m_V, z_L)  
\frac{d^2 \tilde h_V}{d \theta^2}\bigg]_{\theta = \theta_H}  
= \frac{(4\pi)^2 m_H^2 f_H^2}{(kz_L^{-1})^4} ~.
\label{universality3}
\end{align}
We do this procedure for two sets; $(n_F, n_V, c_V) = (n_F^{(1)}, n_V^{(1)}, c_V^{(1)})$ and
$(n_F^{(2)}, n_V^{(2)}, c_V^{(2)})$.  Then (\ref{universality3}) implies that
\begin{align}
&n_F^{(1)} \alpha_F (c_F^{(1)}, z_L) =  n_F^{(2)} \alpha_F (c_F^{(2)}, z_L) \equiv \beta_F ~,  \cr
\noalign{\kern 5pt}
&n_V^{(1)}  \alpha_V (c_V^{(1)}, \tilde m_V^{(1)}, z_L) = n_V^{(2)}  \alpha_V (c_V^{(2)}, \tilde m_V^{(2)}, z_L) 
 \equiv \beta_V~.
\label{universality4}
\end{align}
Although values of $(c_F, \tilde m_V)$ depend on the choice of $(n_F, n_V, c_V)$, 
$\beta_F = n_F \alpha_F (c_F, z_L)$ and $\beta_V = n_V  \alpha_V (c_V, \tilde m_V, z_L)$ are
universal, provided solutions exist.
Consequently one obtains
\begin{align}
&V_\eff (\theta) \simeq \frac{(k z_L^{-1})^4}{(4\pi)^2} \, \bar h (\theta) ~, ~~~
\bar h (\theta) = 
h_0  (\theta) + \beta_F \tilde h_F (\theta)  + \beta_V \tilde h_V (\theta)   ~.
\label{universality5}
\end{align}
It immediately follows that 
\begin{align}
\lambda_3 (\theta_H) &= \frac{g_w m_H^2 \sin\theta_H}{12 m_W} 
\, \frac{\bar h^{(3)} (\theta_H)}{\bar h^{(2)} (\theta_H)} ~, \cr
\noalign{\kern 5pt}
\lambda_4 (\theta_H) &= \frac{g_w^2 m_H^2 \sin^2 \theta_H}{96 m_W^2} 
\, \frac{\bar h^{(4)} (\theta_H)}{\bar h^{(2)} (\theta_H)} ~,
\label{univdersality6}
\end{align}
which explains the $\theta_H$ universality observed in the previous section.
The relevant quantities for $\lambda_3$ and $\lambda_4$ are $\beta_F (\theta_H)$ and 
$\beta_V (\theta_H)$, but not $(n_F, n_V, c_V)$.
As mentioned above, the $z_L$-dependnce of $h_F (\theta ; c_F, z_L)$ and 
$h_V (\theta ; c_V, \tilde m_V, z_L)$ is weak.  
The $\theta_H$ universality stays valid to good approximation even for varying $z_L$.  
For instance,  for $\theta_H=0.1$ and $(n_F, n_V, c_V) = (2, 2, 0)$, 
the resultant $(\lambda_3, \lambda_4)$ is
$(28.93 \,{\rm GeV}, 0.02042)$ for $z_L = 1.237 \times 10^8$, which should be
compared to $(29.03 \,{\rm GeV}, 0.02083)$ for $z_L = 10^{10}$.

The $\theta_H$ universality is observed in other physical quantities. 
The Higgs boson couplings $g_{WWH}$ and $g_{ZZH}$ to $W$, $Z$, 
and Yukawa couplings $y_f$ to quarks and leptons are given, to good approximation,  
by \cite{FHHOY2019b, HK2009}
\begin{align}
&g_{WWH}  = g_w m_W \cos \theta_H ~, \cr
\noalign{\kern 5pt}
&g_{ZZH}  = \frac{g_w m_Z}{\cos \theta_W^0} \, \cos \theta_H ~, \cr
\noalign{\kern 5pt}
&y_f = \begin{cases}  \myfrac{m_f }{v_\SM} \, \cos\theta_H &\hbox{in the A model} \cr
\noalign{\kern 5pt}
 \myfrac{m_f }{v_\SM}  \, \cos^2  \onehalf \theta_H &\hbox{in the B model}   \end{cases} 
\label{Higgscouplings1}
\end{align}
where $v_\SM= f_H \sin\theta_H = 2 m_W/g_w$. For small $\theta_H$, deviation
in the Higgs couplings in (\ref{Higgscouplings1}) is small, whereas deviation in 
$\lambda_3$ and $\lambda_4$ becomes substantial.
We remark that the relations in (\ref{Higgscouplings1}) have been derived in the composite Higgs model 
where the parameter $\sqrt{\xi} = v/f$ corresponds to $\theta_H$ in GHU.\cite{Giudice2007, compositeHiggsRev}

\section{Dark Fermions}

Although the $\theta_H$ universality holds for various couplings associated with the Higgs
boson, masses of dark fermions $\Psi_F$ and $\Psi_{(1, 5)}^\pm$, for instance, 
sensitively depend on the choice of  the parameters $(n_F, n_V, c_V)$.
They are determined by (\ref{functionAdarkF}) for $\Psi_F$,
and by (\ref{functionAdarkV1}) and (\ref{functionAdarkV2}) 
for charged and neutral components of $\Psi^\pm_{(1,5)}$.
In Table \ref{tab:DFmass} their masses 
are tabulated for various $\theta_H$ with $n_F = n_V = 2$.
Dark fermions have relatively small masses compared with the KK mass scale $m_\KK$.
The lightest neutral component of the dark fermions can be a candidate for dark matter.

\begin{table}[tbh]
\renewcommand{\arraystretch}{1.2}
\centering
\caption{
KK mass and dark fermion masses are shown in the unit of TeV  for various $\theta_H$ 
with $z_L = 10^{10}$, $n_F = n_V = 2$ and $c_V=0.2$.  Charged and neutral components of
$\Psi_{(1, 5)}^\pm$ have nearly the same masses.
}
\vskip 10pt
\begin{tabular}{|c||c|c|c|c|}
\hline
$\theta_H$ & $m_{KK}$ & $\Psi_F$ & $\Psi_{(1, 5)}^\pm$   \\
\hline
0.05 & 24.1 & 6.30 & 5.60   \\ 
\hline
0.10 & 12.1 & 3.42 & 2.84   \\ 
\hline
0.15 & 8.07 & 2.38 & 1.91   \\ 
\hline
0.20 & 6.08 & 1.82 & 1.45   \\ 
\hline
\end{tabular}
\label{tab:DFmass}
\end{table}

\begin{figure}[tbh]
\begin{center}
\includegraphics[width=100mm]{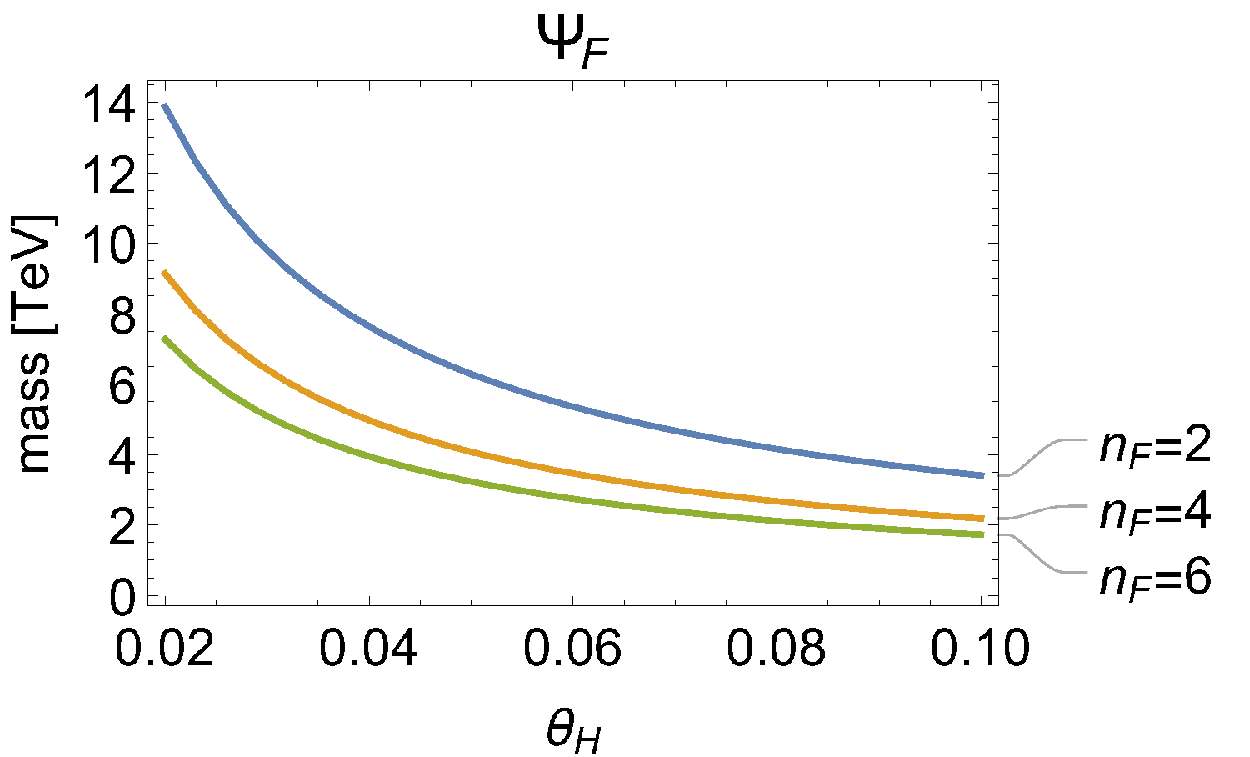}
\caption{$\theta_H$ and $n_F$ dependence of the mass of dark fermion $\Psi_F$.
$z_L = 10^{10}$, $n_V=2, c_V = 0.2$.}   
\label{fig:mass}
\end{center}
\label{fig:DFmass}
\end{figure}

In Fig.\ \ref{fig:DFmass} the mass of $\Psi_F$ is plotted as a function of $\theta_H$
for several $n_F$.    The mass decreases as $n_F$ increases.
Similar behavior is obtained for $\Psi_{(1, 5)}^\pm$ as $n_V$ is varied with $c_V$ fixed.

\section{Summary}

In this paper we have examined the effective potential $V_\eff (\theta_H)$ in 
GUT inspired $SO(5) \times U(1) \times SU(3)$ gauge-Higgs unification to confirm 
that electroweak symmetry breaking is dynamically induced by the Hosotani mechanism.
From $V_\eff (\theta_H)$ the cubic and quartic self-couplings, $\lambda_3$ and $\lambda_4$, 
of the Higgs boson are determined.   We have shown the $\theta_H$ universality of these
couplings, i.e.\ they are determined as functions of $\theta_H$ to high accuracy, 
irrespective of the details of other parameters in the theory.
For $\theta_H=0.1$ ($0.15$),  $\lambda_3$ and $\lambda_4$ are smaller than those
in the standard model by 7.7\% (8.1\%)  and 30\% (32\%),  respectively.
The $\theta_H$ universality in $\lambda_3$ and $\lambda_4$  is understood as a result 
of the factorization property of each component in the contributions to the effective potential, 
which is valid  to high accuracy in the Randall-Sundrum warped space with $z_L \gg 1$.  

The $\theta_H$ universality gives the model great  prediction power.
Once the value of $\theta_H$ is determined by one of the experimental data, then many
other physical quantities such as masses and couplings of various particles are predicted.
It has been known that gauge-Higgs unification models in the RS space predict
large parity violation in the couplings of quarks and leptons to $Z'$ particles (KK  modes of 
$\gamma$, $Z$ and $Z_R$).  Its effect can be clearly seen in electron-positron  collision 
experiments with polarized electron/positron beams 
in which $\theta_H$ is the most important parameter. 
$Z'$ particles can be directly produced at LHC, and parity-violating couplings
would manifest in the rapidity distribution in $t \bar t$ production. 
CKM mixing with natural FCNC suppresion is also incorporated  in the GUT inspired
gauge-Higgs unification.   It is curious to pin down the behavior of the model 
at finite temperature and implications to cosmology.
$SO(5) \times U(1) \times SU(3)$ gauge-Higgs unification is one of the most
promising scenarios beyond the standard model.
We shall come back to these issues in future.

\section*{Acknowledgements}

This work was supported in part 
by European Regional Development Fund-Project Engineering Applications of 
Microworld Physics (No.\ CZ.02.1.01/0.0/0.0/16\_019/0000766) (Y.O.), 
by the National Natural Science Foundation of China (Grant Nos.~11775092, 
11675061, 11521064, 11435003 and 11947213) (S.F.), 
by the International Postdoctoral Exchange Fellowship Program (IPEFP) (S.F.), 
and by Japan Society for the Promotion of Science, 
Grants-in-Aid  for Scientific Research,  No.\ 19K03873 (Y.H.) and Nos.\  18H05543 
and 19K23440 (N.Y.).

\appendix
\section{Mass spectrum}

In evaluating the effective potential $V_\eff (\theta_H)$ in Section  3, one needs to know
the mass spectrum of each KK tower of the fields in the model.  It is sufficient to know
the form of  functions whose zeros determine the mass spectrum.
These functions have been given in ref.\ \cite{FHHOY2019a}.
We summarize them in this appendix for the convenience.

We first introduce 
\begin{eqnarray}
 F_{\alpha, \beta}(u, v) \equiv 
J_\alpha(u) Y_\beta(v) - Y_\alpha(u) J_\beta(v) ~, 
\label{functionA1}
\end{eqnarray}
where $J_\alpha (u)$ and $Y_\alpha (u)$ are the first and second kind Bessel functions.
For gauge fields we define 
\begin{align}
 C(z; \lambda) &= \frac{\pi}{2} \lambda z z_L F_{1,0}(\lambda z, \lambda z_L), 
\nonumber\\
 S(z; \lambda) &= -\frac{\pi}{2} \lambda  z F_{1,1}(\lambda z, \lambda z_L),
\nonumber\\
 C^\prime (z; \lambda) &= \frac{\pi}{2} \lambda^2 z z_L F_{0,0}(\lambda z, \lambda z_L), 
\nonumber\\
S^\prime (z; \lambda) &= -\frac{\pi}{2} \lambda^2 z  F_{1,1}(\lambda z, \lambda z_L). 
\label{functionA2}
\end{align}
For fermion fields with a bulk mass parameter $c$, we define 
\begin{align}
\begin{pmatrix} C_L \cr S_L \end{pmatrix} (z; \lambda,c)
&= \pm \frac{\pi}{2} \lambda \sqrt{z z_L} F_{c+\frac12, c\mp\frac12}(\lambda z, \lambda z_L) ~, \cr
\begin{pmatrix} C_R \cr S_R \end{pmatrix} (z; \lambda,c)
&= \mp \frac{\pi}{2} \lambda \sqrt{z z_L} F_{c- \frac12, c\pm\frac12}(\lambda z, \lambda z_L) ~,
\label{functionA3}
\end{align}
and
\begin{align}
{\cal C}_{R1}(z; \lambda, c, \tilde m) &= C_R(z; \lambda, c +\tilde{m})+C_R(z; \lambda, c -\tilde{m}), 
\nonumber\\ 
{\cal C}_{R2}(z; \lambda, c , \tilde m) &= S_R(z; \lambda, c +\tilde{m})-S_R(z; \lambda,c -\tilde{m}), 
\nonumber\\ 
{\cal S}_{L1}(z; \lambda, c , \tilde m) &= S_L(z; \lambda, c +\tilde{m})+S_L(z; \lambda,c -\tilde{m}), 
\nonumber\\ 
{\cal S}_{L2}(z; \lambda, c , \tilde m) &= C_L(z; \lambda, c +\tilde{m})-C_L(z; \lambda, c -\tilde{m}), 
\nonumber\\ 
{\cal C}_{L1}(z; \lambda, c , \tilde m) &= C_L(z; \lambda, c +\tilde{m})+C_L(z; \lambda, c -\tilde{m}), 
\nonumber\\ 
{\cal C}_{L2}(z; \lambda, c , \tilde m) &= S_L(z; \lambda, c +\tilde{m})-S_L(z; \lambda, c -\tilde{m}), 
\nonumber\\ 
{\cal S}_{R1}(z; \lambda, c , \tilde m) &= S_R(z; \lambda, c +\tilde{m})+S_R(z; \lambda, c -\tilde{m}), 
\nonumber\\ 
{\cal S}_{R2}(z; \lambda, c , \tilde m) &= C_R(z; \lambda, c +\tilde{m})-C_R(z; \lambda, c -\tilde{m}).
\label{functionA4}
\end{align}

\subsection{Gauge bosons}

 The mass spectrum $\{ m_n = k \lambda_n \}$ of $W$ and $W_R$ towers is determined by 
\begin{eqnarray}
 W \ {\rm tower}&:& 
	2 S(1; \lambda)C^\prime(1; \lambda) + \lambda \sin^2 \theta_H = 0,  \cr
 W_R \ {\rm tower}&:& 	C(1; \lambda) = 0. 
 \label{functionA5}
\end{eqnarray}
The spectrum of  $\gamma$, $Z$, $Z_R$ and $A_z$ towers is determined by 
\begin{eqnarray}
 \gamma \ {\rm tower} &:& C^\prime(1; \lambda) = 0, \cr
 Z \ {\rm tower}&:&  
	2 S(1; \lambda)C^\prime(1; \lambda) + (1+s_\phi^2)\lambda \sin^2 \theta_H = 0,  \cr
 Z_R \ {\rm tower}&:& 	C(1; \lambda) = 0 , \cr
 A_z \ {\rm tower}&:& 	S(1; \lambda) C^\prime(1; \lambda) + \lambda \sin^2 \theta_H = 0.  
 \label{functionA6}
 \end{eqnarray}

\subsection{Fermions}
With given up-type quark   masses $m_Q = (m_u, m_c, m_t)$ 
the bulk mass parameter $c_Q= (c_u, c_c, c_t)$ of  up-type quark multiplets is fixed  by 
\begin{eqnarray}
 S_L(1; \lambda, c_Q) S_R(1; \lambda, c_Q) + \sin^2\frac{\theta_H}{2} = 0,  
\label{functionAtop}
\end{eqnarray}
where $\lambda  = \lambda_Q  =m_Q/k$.   
Then the spectrum of up-type quark towers is determined by (\ref{functionAtop}).
In the down-type quark sector there are brane interactions which mix $d^{\prime \alpha}$
and $D^{+ \beta}$ through (\ref{branemass1}).   When  brane interactions are
diagonal in the generation space, $\mu^{\alpha \beta} = \delta^{\alpha \beta} \mu_\alpha$, 
the spectrum of down-type quark tower is determined by
\begin{align}
&\Big( S_L^Q  S_R^Q + \sin^2\frac{\theta_H}{2} \Big)
\big({\cal S}_{L1}^{D}{\cal S}_{R1}^{D}
 -{\cal S}_{L2}^{D}{\cal S}_{R2}^{D}\big) \cr
\noalign{\kern 5pt}
&\hskip 3.cm
+|\mu_1|^2 C_R^Q S_R^Q
\big({\cal S}_{L1}^{D}{\cal C}_{L1}^{D} -{\cal S}_{L2}^{D}{\cal C}_{L2}^{D} \big)=0 ~, \cr
\noalign{\kern 5pt}
&S_L^Q = S_L(1; \lambda, c_Q) , ~~ {\cal S}_{L1}^{D} = {\cal S}_{L1} (1; \lambda, c_D, \tilde m_D) ,
~~ {\rm etc.}
\label{functionAbottom}
\end{align}
where $\mu_1 = (\mu_d, \mu_s, \mu_b)$,  $c_D = (c_{D_d}, c_{D_s}, c_{D_b})$ and
$\tilde m_D = (\tilde m_{D_d}, \tilde m_{D_s}, \tilde m_{D_b})$.
The parameters $\mu_1, c_D, \tilde m_D$ are determined such that 
$\lambda = (\lambda_d, \lambda_s, \lambda_b) = k^{-1} (m_d, m_s, m_b)$
solves (\ref{functionAbottom}) in each generation.
For the third generation, for instance, we take 
$(\mu_b, c_{D_b}, \tilde m_{D_d}) = (0.1, 1.044, 1.0)$.
Only top quark multiplet among quark multiplets gives a relevant contribution to $V_\eff (\theta_H)$.
By considering general $\mu^{\alpha \beta}$ the CKM mixing is incorporated with natural FCNC
suppression.\cite{FHHOY2019b}

With given charged lepton masses $m_L = (m_e, m_\mu, m_\tau)$
the bulk mass parameter $c_L= (c_e, c_\mu, c_\tau)$ of  charged lepton multiplets is fixed  by 
\begin{eqnarray}
 S_L(1; \lambda, c_L) S_R(1; \lambda, c_L) + \sin^2\frac{\theta_H}{2} = 0,  
\label{functionAcharged}
\end{eqnarray}
where $\lambda  = \lambda_L  =m_L/k$.   
Then the spectrum of charged lepton towers is determined by (\ref{functionAcharged}).
In the neutrino sector brane interactions mix $\nu^\alpha, \nu^{\prime \alpha}, \chi^\alpha$.
When both $M^{\alpha\beta} = \delta^{\alpha\beta} M_\alpha$ in (\ref{brane-chi}) and 
$m_B^{\alpha\beta}  = \delta^{\alpha\beta} m_B^\alpha$ in (\ref{branemass2})
are diagonal, the spectrum of neutrino tower is determined by
\begin{align}
&(k\lambda -M) \Big( S_L^L S_R^L + \sin^2 \frac{\theta_H}{2} \Big) 
+ \frac{m_B^2}{k} \, S_R^L C_R^L = 0 ~, \cr
\noalign{\kern 5pt}
&S_R^L = S_R (1; \lambda, c_L) , ~~ {\rm etc.}
\label{functionAneutrino}
\end{align}
where $M = (M_1, M_2, M_3)$ and  $m_B = (m_B^1, m_B^2, m_B^3)$.
With $c_L < - \onehalf$ the light neutrino mass is given by
\begin{align}
&m_\nu \sim \frac{m_L^2 M}{(2|c_L|-1) \, m_B^2}
\label{neutrinomass}
\end{align}
in each generation.  Contributions from lepton multiplets to $V_\eff (\theta_H)$ are
negligible.

The spectrum of  dark fermion $\Psi_F$ tower is determined by 
\begin{eqnarray}
 S_L(1; \lambda, c_F) S_R(1; \lambda, c_F) + \cos^2\frac{\theta_H}{2} = 0 ~.   
\label{functionAdarkF}
\end{eqnarray}
The spectrum of 
charged components of dark fermions $\Psi_{(1,5)}^\pm$ tower is determined by 
\begin{eqnarray}
 {\cal S}_{L1} (1; \lambda, c_V) {\cal S}_{R1} (1; \lambda, c_V)
 - {\cal S}_{L2} (1; \lambda, c_V) {\cal S}_{R2} (1; \lambda, c_V)
 = 0,  
 \label{functionAdarkV1}
\end{eqnarray}
whereas the spectrum of  neutral component tower is determined by 
\begin{align}
&\big\{ B_0(\lambda, c_V, \tilde m_V) - 2 \cos 2 \theta_H \big\}^2=0 ~,  \cr
\noalign{\kern 5pt}
&B_0(\lambda, c, \tilde m) = 
C_L(1; \lambda, c + \tilde m) C_R(1; \lambda, c - \tilde m) 
  + C_L(1; \lambda, c - \tilde m) C_R(1; \lambda, c + \tilde m) \cr
&\quad
  + S_L(1; \lambda, c + \tilde m) S_R(1; \lambda, c - \tilde m) 
  + S_L(1; \lambda, c - \tilde m) S_R(1; \lambda, c + \tilde m).
 \label{functionAdarkV2}
\end{align}
There are two degenerate towers.\footnote{There was a typo in  (D.16) of 
ref.\ \cite{FHHOY2019a}.  The last term in the second line,  
$s_H^2c_H^2 ({\cal C}_{R1}^V{\cal C}_{L2}^V -{\cal C}_{R2}^V{\cal S}_{L1}^V )^2$, 
should be $s_H^2c_H^2 ({\cal C}_{R1}^V{\cal S}_{L2}^V -{\cal C}_{R2}^V{\cal S}_{L1}^V )^2$.
With this correction (D.16) of ref.\ \cite{FHHOY2019a} coincides with 
(\ref{functionAdarkV2}) in the current paper.}

\section{Useful functions}

As shown in the formula (\ref{effVgeneral2}), a mass-determining function $\rho(m; \theta_H)$
is analytically continued to $\rho(iy; \theta_H)$.  We summarize   functions used in the evaluation
of $V_\eff (\theta_H)$ in Section 3.   We introduce
\begin{eqnarray}
 \hat{F}_{\alpha, \beta}(u, v) \equiv 
  I_\alpha(u) K_\beta(v) -e^{-i (\alpha-\beta)\pi} K_\alpha(u) I_\beta(v) ~,
 \label{functionB1}
\end{eqnarray}
where $I_\alpha (u)$ and $K_\alpha (u)$ are first and second kind modified Bessel functions.
In terms of $ \hat{F}_{\alpha, \beta}(u, v)$ we define
\begin{align}
\hat{C}(q) &= q \hat{F}_{1,0}(q z_L^{-1}, q) ~, \cr
\hat{S}(q) &=  i q z_L^{-1} \hat{F}_{1,1}(q z_L^{-1}, q) ~, \cr
\hat{C}^\prime (q) &= q^2 z_L^{-1} \hat{F}_{0,0}(q z_L^{-1}, q) ~,  \cr 
\hat{S}^\prime (q) &= - i q^2 z_L^{-2} \hat{F}_{1,1}(q z_L^{-1}, q) 
\label{functionB2}
\end{align}
for gauge fields.
For fermion fields with $c>0$ we define
\begin{align}
\hat{C}_L(q; c) &= q z_L^{-1/2}\hat{F}_{c+\frac12, c-\frac12}(q z_L^{-1}, q) ~, \cr
\hat{S}_L(q; c) &= i q z_L^{-1/2}\hat{F}_{c+\frac12, c+\frac12}(q z_L^{-1}, q) ~, \cr
\hat{C}_R(q; c) &= q z_L^{-1/2}\hat{F}_{c-\frac12, c+\frac12}(q z_L^{-1}, q) ~, \cr
\hat{S}_R(q; c) &= - i q z_L^{-1/2}\hat{F}_{c-\frac12, c-\frac12}(q z_L^{-1}, q) ~.
\label{cshat} 
\end{align}
For $c<0$, we use the  the relations  
\begin{eqnarray}
 \hat{C}_L(q; -c) =   \hat{C}_R(q; c) ~, ~~
 \hat{S}_L(q; -c) = - \hat{S}_R(q; c) ~.
\label{cshat2} 
\end{eqnarray}


\vskip 1.cm

\def\jnl#1#2#3#4{{#1}{\bf #2},  #3 (#4)}

\def\Zphys{{\em Z.\ Phys.} }
\def\jssc{{\em J.\ Solid State Chem.\ }}
\def\jpsJ{{\em J.\ Phys.\ Soc.\ Japan }}
\def\ptps{{\em Prog.\ Theoret.\ Phys.\ Suppl.\ }}
\def\PTP{{\em Prog.\ Theoret.\ Phys.\  }}
\def\PTEP{{\em Prog.\ Theoret.\ Exp.\  Phys.\  }}
\def\JMP{{\em J. Math.\ Phys.} }
\def\NPB{{\em Nucl.\ Phys.} B}
\def\NP{{\em Nucl.\ Phys.} }
\def\PLB{{\it Phys.\ Lett.} B}
\def\PL{{\em Phys.\ Lett.} }
\def\PRL{\em Phys.\ Rev.\ Lett. }
\def\PRB{{\em Phys.\ Rev.} B}
\def\PRD{{\em Phys.\ Rev.} D}
\def\PRe{{\em Phys.\ Rep.} }
\def\AP{{\em Ann.\ Phys.\ (N.Y.)} }
\def\RMP{{\em Rev.\ Mod.\ Phys.} }
\def\ZPC{{\em Z.\ Phys.} C}
\def\SCI{\em Science}
\def\CMP{\em Comm.\ Math.\ Phys. }
\def\MPLA{{\em Mod.\ Phys.\ Lett.} A}
\def\IJMPA{{\em Int.\ J.\ Mod.\ Phys.} A}
\def\IJMPB{{\em Int.\ J.\ Mod.\ Phys.} B}
\def\EPJC{{\em Eur.\ Phys.\ J.} C}
\def\PR{{\em Phys.\ Rev.} }
\def\JHEP{{\em JHEP} }
\def\JCAP{{\em JCAP} }
\def\cmp{{\em Com.\ Math.\ Phys.}}
\def\JPA{{\em J.\  Phys.} A}
\def\JPG{{\em J.\  Phys.} G}
\def\NJP{{\em New.\ J.\  Phys.} }
\def\CQG{\em Class.\ Quant.\ Grav. }
\def\ATMP{{\em Adv.\ Theoret.\ Math.\ Phys.} }
\def\ibid{{\em ibid.} }
\def\ChP{{\em Chin.Phys.}C}


\renewenvironment{thebibliography}[1]
         {\begin{list}{[$\,$\arabic{enumi}$\,$]}  
         {\usecounter{enumi}\setlength{\parsep}{0pt}
          \setlength{\itemsep}{0pt}  \renewcommand{\baselinestretch}{1.2}
          \settowidth
         {\labelwidth}{#1 ~ ~}\sloppy}}{\end{list}}

\end{document}